\documentclass[12pt, reqno]{amsart}

\IfFileExists{mymtpro2.sty}{%
  \usepackage[subscriptcorrection]{mymtpro2}
}{}

\usepackage{a4, upgreek, amssymb, amsmath}
\usepackage{appendix}

\newtheorem{theorem}{Theorem}[section]
\newtheorem{lemma}{Lemma}[section]
\newtheorem{corollary}{Corollary}[section]
\newtheorem{prop}{Proposition}[section]
\theoremstyle{definition}

\newtheorem{remarks}[theorem]{Remarks}
\newtheorem{example}{Example}[section]

\newcommand{\labelnummer}{\mbox{\normalfont (\roman{numcount})}}%

\makeatletter

  {\let\curlabelspeicher\@currentlabel%
    \begin{list}{\labelnummer}%
      {\usecounter{numcount}\leftmargin0pt%
        \topsep0.5ex\partopsep2ex\parsep0pt\itemsep0ex\@plus1\p@%
        \labelwidth2.5em\itemindent3.5em\labelsep1em%
      }%
    \let\saveitem\item%
    \def\item{\saveitem%
      \def\@currentlabel{{\upshape\curlabelspeicher}$\,$\labelnummer}}%
    \let\savelabel\label%
    \def\label##1{\savelabel{##1}%
      \@bsphack%
        \ifmmode\else%
          \protected@write\@auxout{}%
          {\string\newlabel{##1item}{{\labelnummer}{\thepage}}}%
        \fi%
      \@esphack%
    }%
  }{\end{list}}%

\renewcommand{\appendix}{\def\thesection{\textsc{Appendix}}}

 \let\leq\le
 \let\geq\ge

\DeclareMathOperator{\tr}{tr\kern1pt}

\newcommand\ZZ{\mathbb Z}

\makeatletter

\newif\ifper\pertrue
\def\per{.}

\def\bti{\@ifnextchar[\bbti\bbbti}
\def\bbti[#1]#2{#2, #1.}
\def\bbbti#1{#1.}

\def\z{\@ifnextchar[\zz\zzz}
\def\zz[#1]#2#3#4#5{\perfalse\emph{#2} \textbf{#3}, #4 (#5) [#1]}
\def\zzz#1#2#3#4{\emph{#1} \textbf{#2}, #3 (#4)\ifper\per\fi\pertrue}

\def\pub{\@ifstar\pubstar\pubnostar}
\def\pubnostar{\@ifnextchar[\@@pubnostar\@pubnostar}
\def\@@pubnostar[#1]#2#3#4{#2, #3, #4, #1\ifper\per\fi\pertrue}
\def\@pubnostar#1#2#3{#1, #2, #3\ifper\per\fi\pertrue}
\def\pubstar[#1]#2#3#4{\perfalse #2, #3, #4 [#1]\pertrue}

\makeatother

\newcommand{\bel}{\begin{equation} \label}
\newcommand{\ee}{\end{equation}}

\def\beq{\begin{equation}}
\def\eeq{\end{equation}}
\newcommand{\bea}{\begin{eqnarray}}
\newcommand{\eea}{\end{eqnarray}}
\newcommand{\beas}{\begin{eqnarray*}}
\newcommand{\eeas}{\end{eqnarray*}}

{

\newcommand{\Pp}{\mathbb{P}}
\newcommand{\R}{\mathbb{R}}

\newcommand{\Z}{\mathbb{Z}}

\newcommand{\N}{\mathbb{N}}

\newcommand{\C}{\mathbb{C}}
\newcommand{\E}{\mathbb{E}}

\begin{document}

\title[Density of states outer measure]{Dependence of the density of states outer measure on the potential for deterministic Schr\"odinger operators on graphs with applications to ergodic and random models}

\author[P.\ D.\ Hislop]{Peter D.\ Hislop}
\address{Department of Mathematics,
    University of Kentucky,
    Lexington, Kentucky  40506-0027, USA}
\email{peter.hislop@uky.edu}

\author[C.\ A.\ Marx]{Christoph A.\ Marx}
\address{Department of Mathematics,
Oberlin College,
Oberlin, Ohio 44074, USA}
\email{cmarx@oberlin.edu}


\begin{abstract}
We continue our study of the dependence of the density of states measure and related spectral functions of 
Schr\"odinger operators on the potential. Whereas our earlier work focused on random Schr\"odinger operators,
 we extend these results to Schr\"odinger operators on infinite graphs with deterministic potentials and ergodic potentials, and improve our results for random potentials. In particular, for random Schr\"odinger operators on the lattice, we prove the Lipschitz continuity of the DOSm in the single-site measure, recovering results of \cite{kachkovskiy, shamis}. For our treatment of deterministic potentials, we first study the density of states outer measure (DOSoM), defined for all Schr\"odinger operators, and prove a deterministic result of the modulus of continuity of the DOSoM with respect to the potential. We apply these results to Schr\"odinger operators on the lattice $\Z^d$ and the Bethe lattice. In the former case, we prove the Lipschitz continuity of the DOSoM, and in the latter case, we prove that the DOSoM is $\frac{1}{2}$-log-H\"older continuous. Our technique combines the abstract Lipschitz property of one-parameter families of self-adjoint operators with a new finite-range reduction that allows us to study the dependency of the DOSoM and related functions on only finitely-many variables and captures the geometry of the graph at infinity.    
\end{abstract}

\maketitle \thispagestyle{empty}

\tableofcontents

\vspace{.2in}



\section{Introduction}\label{sec:introduction1}
\setcounter{equation}{0}

This is the third of a series of papers \cite{hislop_marx_1, hislop_marx_2} in which we explore the dependence of the density of states measure (DOSm) for Schr\"odinger operators on the potential. In this article, we extend and refine previous results by considering the density of states outer measure (DOSoM), defined in section \ref{subsec:DOSoMdefn1}. The DOSoM is a generalization of the DOSm defined for all deterministic Schr\"odinger operators. It was first introduced in \cite{bourgain-klein1}  by Bourgain and Klein for Schr\"odinger operators on $\Z^d$ and $\R^d$, and we extend this definition to Schr\"odinger operators on infinite graphs.  
The DOSoM is a deterministic quantity which allows us to study its dependence on the potential sequence for discrete Schr\"odinger operators on graphs. 

We prove that the DOSoM is continuous with respect to the potential in the $\ell^\infty$-norm, and we obtain an explicit bound on the modulus of continuity in terms of the uniform growth function describing the underlying graph. Our proof uses a deterministic formulation of the finite-volume reduction from \cite{hislop_marx_1} and the Lipschitz property established in \cite{hislop_marx_2}. With these constructions, we obtain optimal results on the modulus of continuity for lattices. For the Bethe lattice, we prove that the DOSoM is $\frac{1}{2}$-log-H\"older continuous. 
We also obtain quantitative bounds on the modulus of continuity for the integrated outer density of states (IoDS), the cumulative distribution of the DOSoM. 


In addition, we discuss the implications of these new deterministic results for general ergodic Schr\"odinger operators  on lattices and more general infinite graphs. We also apply our results to random Schr\"odinger operators by treating them as a subset of  ergodic Schr\"odinger operators through the use of the quantile function associated with the single-site probability measure. 
For random Schr\"odinger operators on lattices, these new deterministic results allow us to improve the known quantitative estimates on the modulus of continuity of the DOSm and the IDS with respect to the single-site probability measure.
We also obtain new and improved results on the modulus of continuity of the DOSm and IDS with respect to the disorder in the weak coupling limit, previously obtained in \cite{hks, schenker} and in \cite{hislop_marx_1}.

For a more detailed introduction to the problem of the dependence of the DOSm on the potential, motivation,  and historical commentary, we refer the reader to the introductions of \cite{hislop_marx_1, hislop_marx_2}

After we completed our work, I.\  Kachkovskiy \cite{kachkovskiy} told us about the work of Aleksandrov and Peller \cite{aleksandrovPeller2011}, discussed in section \ref{app:AltProofAP1}. Their general functional analytic estimate, presented as Theorem \ref{thm:peller1}, provides for an improvement of our result, Theorem \ref{thm_Bethe}, for the Bethe lattice.
For lattices, however, their result leads to a weaker bound than our optimal result, Theorem \ref{thm_lattice_DOSoM}.
We refer the reader to section \ref{app:AltProofAP1} for more information. 

A.\ Skripka also pointed out to us the similarity between some of her results, in particular \cite[Lemma 3.8(3)]{skripka2011},
and the Lipschitz property in Proposition \ref{prop_lipschitz_traceclass}.


\subsection{Contents}\label{subsec:contents1}

In section \ref{sec:DOSoM1}, we describe the family of infinite graphs with a uniform growth hypothesis (UGH) to which our results apply. The UGH quantifies the ratio of surface area to volume of regions in the graph as the region tends to the graph. 
We define Schr\"odinger operators on these graphs and the density of states outer measure by extending the definition of Bourgain and Klein \cite{bourgain-klein1} from lattices to graphs. The main result on the modulus of continuity of the DOSoM for Schr\"odinger operators is presented and proved in section \ref{sec:quantDOSoM1}. Our first applications to deterministic Schr\"odinger operators are presented in section \ref{sec:applications1}, and our second set of applications to ergodic and random Schr\"odinger operators appears in section \ref{sec:applications2}.  The paper concludes with four appendices.
In Appendix \ref{app_DOSoM_lattice}, we present details of our extension of the Bourgain-Klein \cite{bourgain-klein1} construction of the DOSoM to graphs. In the second, Appendix \ref{app:AltProofAP1}, we discuss the relation between our results and those based on a theorem of Alexsandrov and Peller \cite{aleksandrovPeller2011}.
The third, Appendix \ref{app:equivmetric}, contains a proof of the equivalence of the Fortet-Mourier metric for weak convergence and the Kantorovich-Rubinstein-Wasserstein metric. Finally, in the fourth, Appendix \ref{app:spectrumhausdorff},  we  discuss the continuity of the spectrum with respect to the potential. We prove a theorem on the Hausdorff distance between the a.s.-spectraa of ergodic and random Schr\"odinger operators with respect to the sampling function or the single-site probability measure, respectively. 

\vspace{.2in}
\noindent
\textbf{Acknowledgements:}
We thank I.\ Kachkovskiy for many fruitful discussions and for alerting us to the work of Alexsandrov and Peller \cite{aleksandrovPeller2011}. We also thank A.\ Skripka for discussions on \cite{skripka2011} and related works.


\section{The set-up: Schr\"odinger operators on graphs and the DOSoM}\label{sec:DOSoM1}

The main result of this section, Theorem \ref{thm_DOSoM_main}, applies to discrete Schr\"odinger operators on infinite graphs satisfying the uniform growth hypothesis defined in section \ref{subsec:UGH1}. The modulus of continuity is explicitly expressed in terms of a growth function $\gamma_{\mathbb{G}}$ which captures the geometry at infinity of the graph. As we will show, the choice of this function allows us to prove that the DOSoM is Lipschitz continuous for lattices.


\subsection{Infinite graphs with uniform growth}\label{subsec:UGH1}

In this and the next subsection, we review the construction of Schr\"odinger operators on infinite graphs satisfying certain growth conditions.
Let $\mathbb{G}:=(\mathcal{V},\mathcal{E})$ be an infinite, connected graph with vertices $\mathcal{V}$, edges $\mathcal{E}$, and natural metric $d_\mathbb{G}$ defined as follows: For any two vertices $x,y \in \mathcal{V}$, let $\gamma_{(x,y)}$
be a path in $\mathbb{G}$ connecting $x$ to $y$ and let $N_{\mathbb{E}}(\gamma_{(x,y)})$ be the number of edges comprising the path. Given this set-up, we define a metric by 
\begin{equation}
d_\mathbb{G}(x,y) := ~ \min_{ \{ \gamma_{(x,y)} \}}  N_{\mathbb{E}} ( \gamma_{(x,y)} ) . 
\end{equation}
We will further assume that the graph $\mathbb{G}$ satisfies the following hypothesis:

\vspace{0.1 cm}
\noindent
{\textbf{Uniform growth hypothesis (UGH):}}\emph{ For each $x \in \mathcal{V}$ and $L \in \mathbb{N}$, the closed ball centered at $x$ with radius $L$ ,
\begin{equation} \label{eq_closedballs}
\Lambda_{L}^{(\mathbb{G})}(x): = \{ y \in \mathcal{V} ~:~ d_\mathbb{G}(x,y) \leq L \} ~\mbox{,}
\end{equation}
is a {\em{finite}} set which admits a {\bf{uniform growth function,}} in the sense that there exists a strictly increasing function $\gamma_{\mathbb{G}}: [1, +\infty) \to [1, +\infty)$ such that for each $n \in \mathbb{N}$:
\begin{equation}  \label{eq_growthfunction}
\limsup_{L \to \infty} \left\{ \sup_{x \in \mathcal{V}} \dfrac{\vert \Lambda_{L+n}^{(\mathbb{G})}(x) \vert}{\vert \Lambda_{L}^{(\mathbb{G})}(x) \vert } \right\} \leq \gamma_{\mathbb{G}}(n) ~\mbox{.}
\end{equation}
\vspace{0.1 cm}
Here, and in the following, given a set $A \subseteq \mathcal{V}$, $\vert A \vert$ denotes the number of vertices in $A$. }

In this note, the most important examples for graphs satisfying the UGH are:
\begin{enumerate}
\item The $d$-dimensional lattice $\mathbb{Z}^d$ where 
\begin{align} \label{eq_latticegrowth}
\limsup_{L \to \infty} \left\{ \sup_{x \in \mathcal{V}} \dfrac{\vert \Lambda_{L+n}^{(\mathbb{Z}^d)}(x) \vert}{\vert \Lambda_{L}^{(\mathbb{Z}^d)}(x) \vert } \right\} \leq \limsup_{L \to \infty} \dfrac{(2(L+n) + 1)^d}{(2(L-1) + 1)^d} = 1  ~\mbox{,}
\end{align}
in particular, {\em{any}} strictly increasing function $\gamma_{\mathbb{Z}^d}: [1,+\infty) \to [1,+\infty)$ can serve as a uniform growth function.
\item Other lattices in $\Z^2$, such as hexagonal and trianglular lattices, satisfy the same uniform growth conditions as in \eqref{eq_latticegrowth}. 
\item The Bethe lattice $\mathbb{B}_k$ with coordination number $k \geq 3$ ($k=2$ corresponds to $\mathbb{Z}$). Using the fact that for all $L \in \mathbb{N}$ and $x \in \mathcal{V}$, one has 
\begin{equation}
\vert \Lambda_L^{(\mathbb{B}_k)}(x) \vert = 1 + k \dfrac{(k-1)^{L} - 1}{k - 2} ~\mbox{,}
\end{equation}
we conclude that
\begin{align}  \label{eq_bethegrowth}
\limsup_{L \to \infty} \left\{ \sup_{x \in \mathcal{V}} \dfrac{\vert \Lambda_{L+n}^{(\mathbb{B}_k)}(x) \vert}{\vert \Lambda_{L}^{(\mathbb{B}_k)}(x) \vert } \right\} = \limsup_{L \to \infty} \dfrac{k(k-1)^{L+n} - 2}{ k (k-1)^{L} - 2 } = (k-1)^n := \gamma_{\mathbb{B}_k}(n)   ~\mbox{}
\end{align}
may serve as a uniform growth function.
\end{enumerate}


\subsection{Schr\"odinger operators on graphs}\label{subsec:SchrOpGraph1}

For a real-valued sequence $V \in \ell^\infty(\mathbb{G}; \mathbb{R})$ (``\emph{the potential sequence}''), we consider the Schr\"odinger operator $H_V:~\ell^2(\mathbb{G}; \C) \to \ell^2(\mathbb{G}; \C)$ given by
\begin{equation} \label{eq_Schrodop}
H_V \psi := \Delta_\mathbb{G} + T_V ~\mbox{.}
\end{equation}
Here, $T_V: \ell^2(\mathbb{G}; \C) \to \ell^2(\mathbb{G}; \C)$ denotes the multiplication operator by the sequence $V$,
\begin{equation}
(T_V \psi)(x) = V(x) \psi(x), ~~ \forall x \in \mathcal{V} ~\mbox{,}
\end{equation}
and $\Delta_\mathbb{G}: \ell^2(\mathbb{G}; \C) \to \ell^2(\mathbb{G}; \C)$ is the discrete Laplacian on $\mathbb{G}$,
\begin{equation} \label{eq_laplacian}
(\Delta_\mathbb{G})(x):= \sum_{y \sim x} \psi(y),  ~~ \forall x \in \mathcal{V}  ~\mbox{,}
\end{equation}
where, given a vertex $x \in \mathcal{V}$, the notation $y \sim x$ in (\ref{eq_laplacian}) denotes all the vertices $y \in \mathcal{V}$ which are directly connected to $x$, or equivalently, which satisfy $d_\mathbb{G}(y,x) = 1$. 
The graph Laplacian $\Delta_\mathbb{G}$ is bounded and we write its spectrum as $\sigma (\Delta_\mathbb{G}) = [- \rho_\mathbb{G}, \rho_\mathbb{G}]$, where $\rho_\mathbb{G}$ is the spectral radius of $\Delta_\mathbb{G}$.  


\subsection{Density of states outer measure (DOSoM)}\label{subsec:DOSoMdefn1}

In this note we are interested in quantifying the dependence of the {\em{density of states outer measure}} (DOSoM) of (\ref{eq_Schrodop}) on the potential sequence. The DOSoM serves as a generalization of the ``usual'' density of states measure, which is well defined only for certain models, e.g. periodic or random operators.

The DOSoM was first considered by  Bourgain and Klein in \cite{bourgain-klein1} for Schr\"odinger operators on the lattice $\Z^d$  and the continuum $\R^d$. 
Here we present an appropriate generalization to Schr\"odinger operators on {\em{arbitrary}} graphs. As we will see, allowing for arbitrary graphs will, in general, necessitate working with infinite volume operators as opposed to the finite volume restrictions considered earlier in  \cite{bourgain-klein1} for the lattice case $\mathbb{G} = \mathbb{Z}^d$. To compare with the latter, we note that for $\mathbb{G} = \mathbb{Z}^d$, the metric $d_\mathbb{G}$ is equivalent to the $1$-norm, in particular $\Lambda_L^{(\mathbb{Z}^d)}(x)$ is a closed cube centered at $x$ with vertices on lines parallel to the coordinate axes. 

To define the DOSoM, we denote by $\{\delta_y ~:~ y \in \mathcal{V}\}$ the standard basis in $\ell^2(\mathbb{G}; \C)$ and by
\begin{equation} \label{eq_elementarycoordproj}
\pi_y:=\vert \delta_y \rangle \langle \delta_y \vert ~\mbox{, } y \in \mathcal{V} ~\mbox{,}
\end{equation}
the associated rank-one projections. Moreover, for $L \in \mathbb{N}$, we let
\begin{equation}
P_L^{(\mathbb{G})}(x):= \sum_{y \in \Lambda_L^{(\mathbb{G})}(x)} \pi_y ~\mbox{}
\end{equation}
be the orthogonal projection onto the subspace of $\ell^2(\mathbb{G}; \C)$ associated with the vertices in $\Lambda_L^{(\mathbb{G})}(x)$. Then, we define the DOSoM associated with the operator $H_V$ in (\ref{eq_Schrodop}) by
\begin{equation} \label{eq_defnDOSoM}
n_V^*(f):= \limsup_{L \to \infty} \left\{ \sup_{x \in \mathcal{V}} \dfrac{1}{\vert \Lambda_L^{(\mathbb{G})}(x) \vert} \mathrm{Tr}\left( P_L^{(\mathbb{G})}(x) f(H_V) P_L^{(\mathbb{G})}(x) \right) \right\} ~\mbox{,}
\end{equation}
where $f$ is a bounded (possibly $\mathbb{C}$-valued) Borel function on $\mathbb{R}$. As pointed out earlier, the merit of introducing the DOSoM is that the latter always exists, even for models where the usual density of states measure does not. In addition, we will also obtain results for the {\em{local DOSoM}} at $x \in \mathcal{V}$, defined by
\begin{equation}\label{eq:localDOSoM4}
n_{V;x}^*(f):= \limsup_{L \to \infty} \left\{ \dfrac{1}{\vert \Lambda_L^{(\mathbb{G})}(x) \vert} \mathrm{Tr}\left( P_L^{(\mathbb{G})}(x) f(H_V) P_L^{(\mathbb{G})}(x) \right)  \right\} ~\mbox{,}
\end{equation}
for bounded Borel functions $f$. This local DOSoM will be relevant when we discuss ergodic Schr\"odinger operators. In the ergodic case, the local DOSoM is independent of $x \in \mathcal{V}$ and equals the DOSm almost surely. 


To contrast our definition of the DOSoM in (\ref{eq_defnDOSoM}) with the lattice case considered by Bourgain and Klein in \cite{bourgain-klein1}, we observe that the authors in \cite{bourgain-klein1}  introduced the DOSoM for $\mathbb{G} = \mathbb{Z}^d$ by replacing the ``infinite-volume operator'' $H_V$ in (\ref{eq_defnDOSoM}) by ``finite volume restrictions,'' thus effectively considering
\begin{equation} \label{eq_finitevolumerestr}
H_{V;L}(x):= P_L^{(\mathbb{Z}^d)}(x) H_V P_L^{(\mathbb{Z}^d)}(x) ~\mbox{, for $L > 0$.}
\end{equation}
Appealing to the second resolvent identity shows that for each $a \in \mathbb{C} \setminus \mathbb{R}$ and 
\begin{equation} \label{eq_2ndresolvent_decay}
f_a(t):=(t-a)^{-1}, 
\end{equation}
one has:
\begin{align} 
\left\vert \dfrac{1}{\vert \Lambda_L^{(\mathbb{Z}^d)}(x) \vert} \right. & \left. \mathrm{Tr}\left( P_L^{(\mathbb{Z}^d)}(x) f_a(H_V) P_L^{(\mathbb{Z}^d)}(x) \right) - \dfrac{1}{\vert \Lambda_L^{(\mathbb{Z}^d)}(x) \vert} \mathrm{Tr}\left( f_a( H_{V;L}(x) ) \right) \right\vert \nonumber \\ & \lesssim \sup_{x \in \mathbb{Z}^d} \dfrac{\vert \Lambda_{L+1}^{(\mathbb{Z}^d)}(x) \setminus \Lambda_{L-2}^{(\mathbb{Z}^d)}(x) \vert}{ \vert \Lambda_L^{(\mathbb{Z}^d)}(x) \vert } \lesssim \dfrac{1}{L} ~\mbox{, for all $x \in \mathbb{Z}^d$ ,} \label{eq_compareBKdefn}
\end{align}
which, using a density argument and monotonicity, implies that the definition in (\ref{eq_defnDOSoM}) agrees with the definition in \cite{bourgain-klein1} for the special case that $\mathbb{G} = \mathbb{Z}^d$. There is also a technical issue due to the fact that outer measures are, in general, only subadditive. For completeness, we include the details of the equivalence of our definition \eqref{eq_defnDOSoM} of the DOSoM with the one of Bourgain-Klein in Appendix \ref{app_DOSoM_lattice}. 

In view of lattice Schr\"odinger operators ($\mathbb{G} = \mathbb{Z}^d$), we also note that one more commonly uses closed balls in the $\infty$-norm on $\mathbb{Z}^d$,
\begin{equation}
\Lambda_{L; \infty}^{(\mathbb{Z}^d)}(x):=\{ y \in \mathbb{Z}^d ~:~ \Vert x - y \Vert_\infty \leq L\} ~\mbox{,}
\end{equation}
to define the DOSoM as opposed to the $1$-norm considered in (\ref{eq_closedballs}). Since both these norms are, however, equivalent, i.e.\
\begin{align}
\Lambda_{L-1; \infty}^{(\mathbb{Z}^d)}(x) \leq \Lambda_L^{(\mathbb{Z}^d)}(x) \leq \Lambda_{L; \infty}^{^{(\mathbb{Z}^d)}}(x) ~\mbox{,}
\end{align}
and
\begin{equation}
(2L - 1)^d \leq \vert \Lambda_L^{(\mathbb{Z}^d)}(x) \vert \leq (2L+1)^d ~\mbox{,}
\end{equation}
the definition of the DOSoM is independent of the specific norm used to define balls. 

We emphasize that for more general graphs $\mathbb{G} \neq \mathbb{Z}^d$, the second term on the right hand side of (\ref{eq_compareBKdefn}) may {\em{not}} decay to zero as $L \to \infty$. For instance, for the Bethe lattice ${\mathbb{B}_k}$, the error term is only bounded but not decaying, 
\begin{equation}
\dfrac{\vert \Lambda_{L+1}^{({\mathbb{B}_k})}(x) \setminus \Lambda_{L-2}^{({\mathbb{B}_k})}(x) \vert}{ \vert \Lambda_L^{({\mathbb{B}_k})}(x) \vert } \sim \dfrac{\mathrm{k}^{L+1} - \mathrm{k}^{L-2}}{\mathrm{k}^L} = O(1) ~\mbox{, as }  L \to \infty ~\mbox{.}
\end{equation}
Indeed, this is why for ergodic Schr\"odinger operators on $\mathbb{B}_k$, the density of states measure has to be defined as an analogue of (\ref{eq_defnDOSoM}), using the {\em{infinite volume}} operator $H_V$ instead of the finite volume restrictions in (\ref{eq_finitevolumerestr}); see \cite[Appendix]{AcostaKlein}.


\section{Quantitative continuity of the DOSoM}\label{sec:quantDOSoM1}
\setcounter{equation}{0}

In this section, we prove estimates on the modulus of continuity of the DOSoM for rather general Schr\"odinger operators $H_V$ on infinite graphs satisfying the UGH as defined in \eqref{eq_Schrodop}-\eqref{eq_laplacian}. The modulus of continuity in Theorem \ref{thm_DOSoM_main} is expressed in terms of a uniform growth function $\gamma_{\mathbb{G}}$ for the graph $\mathbb{G}$ and the $L^\infty$-norm of the potential sequence. This main result is based on  the  finite-range reduction presented in section \ref{subsec:FiniteRange1} and the Lipschitz property presented in section \ref{subsec:LipProp1}.

\subsection{Finite-range reduction}\label{subsec:FiniteRange1}

As explained in section \ref{subsec:DOSoM1}, we  must use infinite-volume operators as in (\ref{eq_defnDOSoM}) in order to study the dependence of the DOSoM on the potential sequence.  This 
necessitates the use of  a reductive procedure to finite subgraphs so that we can consider variations of the potential at only finitely-many sites at a time. Similar reductions were carried out in \cite{hislop_marx_1, hislop_marx_2}. 

The latter is accomplished by what will subsequently be referred to as the ``{\em{finite-range reduction}}.'' The key feature underlying this finite range reduction is the following consequence of UGH: For every polynomial $p$, $x \in \mathcal{V}$, and $L \in \mathbb{N}$, the map
\begin{equation} \label{eq_mapfinite_1}
\ell^\infty(\mathbb{G}) \ni \omega \mapsto \mathrm{Tr}\left(  P_L^{(\mathbb{G})}(x) p(H_\omega) P_L^{(\mathbb{G})}(x) \right) \in \mathbb{R}
\end{equation}
only depends on {\em{finitely}} many elements of the sequence $\omega$. Indeed, this map depends at most on the entries $\omega_y$, $y \in \mathcal{V}$, of the sequence $\omega$ which satisfy that 
\begin{equation} \label{eq_mapfinite_2}
y \in \bigcup_{z \in \Lambda_L^{(\mathbb{G})}(x)} \Lambda_{ \lfloor \deg(p)/2 \rfloor }(z) \subseteq \Lambda_{L+  \lfloor \deg(p)/2 \rfloor}(x) ~\mbox{.}
\end{equation}

To formulate the finite range reduction, we introduce the following modification of a given a potential sequence $V \in \ell^\infty(\mathbb{G};\mathbb{R})$: If $x \in \mathcal{V}$, $R>0$, and $W \in \ell^\infty(\mathbb{G}; \R)$, we define the {\em{$(R;W)$-modification of the potential $V$ at $x$}} as the sequence given by
\begin{equation} \label{eq_modifiedpotential}
V_W^{(R;x)}(y):=\begin{cases}  V(y) & ~\mbox{, if } y \in \Lambda_R(x) ~\mbox{, } \\
                                                  W(y) & ~\mbox{, if } y \not \in \Lambda_R(x)  ~\mbox{. }
\end{cases}
\end{equation}

Finally, as in section \eqref{subsec:SchrOpGraph1}, we write $\rho_\mathbb{G}$ for the spectral radius of $\Delta_\mathbb{G}$, in particular for a given a potential $V \in \ell^\infty(\mathbb{G}; \R)$, the spectrum of $H_V$ in (\ref{eq_Schrodop}) satisfies
\begin{equation} 
\sigma(H_V) \subseteq [-\rho_\mathbb{G} - \Vert V \Vert_\infty , \rho_\mathbb{G} + \Vert V \Vert_\infty] ~\mbox{.}
\end{equation}

We are now ready to formulate the finite-range reduction, which we note, is a deterministic version of Lemma 2.1 in \cite{hislop_marx_1}.
 
\begin{lemma}[Finite-range reduction] \label{lemma_finiterange}
Let $f \in \mathcal{C}(\mathbb{R})$, $C>0$, and $\epsilon > 0$ be given. Suppose $p$ is a polynomial satisfying that
\begin{equation} \label{eq_finiterange_hypoth}
\Vert f - p \Vert_{\infty; [-\rho_\mathbb{G} - C , \rho_\mathbb{G} + C]} < \epsilon ~\mbox{.}
\end{equation}
Then, for each $x \in \mathcal{V}$ and $L>0$, letting $M:= L+  \lfloor \deg(p)/2 \rfloor$, one has that for all $V,W \in \ell^\infty(\mathbb{G};[-C,C])$:
\begin{align}
\left\vert \dfrac{1}{\vert \Lambda_L^{(\mathbb{G})}(x) \vert} \left[ \mathrm{Tr} \left( P_L^{(\mathbb{G})}(x) f(H_V) P_L^{(\mathbb{G})}(x) \right) - \mathrm{Tr} \left( P_L^{(\mathbb{G})}(x) f(H_{V_W^{(M;x)}})  P_L^{(\mathbb{G})}(x) \right) \right] \right\vert \leq 2 \epsilon  ~\mbox{.} 
\end{align}
\end{lemma}

\noindent
Here, for a bounded function $g \in \mathcal{C}(\mathbb{R})$, we let $\Vert g \Vert_{\infty;A}$ denote the sup-norm of $g$ on a given set $A \subseteq \mathbb{R}$.

\begin{proof}
Let $V,W \in \ell^\infty (\mathbb{G};[-C,C])$, $L \in \mathbb{N}$, and $x \in \mathcal{V}$.
The key observation underlying the finite range reduction is based on (\ref{eq_mapfinite_1}) and (\ref{eq_mapfinite_2}). This locality allows us to conclude that
\begin{equation} \label{eq_finiterangereduction_key}
\mathrm{Tr} \left( P_L^{(\mathbb{G})}(x) p(H_V) P_L^{(\mathbb{G})}(x) \right) = \mathrm{Tr} \left( P_L^{(\mathbb{G})}(x) p(H_{V_W^{(M;x)}}) P_L^{(\mathbb{G})}(x) \right) ~\mbox{,}
\end{equation}
since the trace in (\ref{eq_finiterangereduction_key}) does not depend on the potential outside of $\Lambda_{M}(x)$.
Moreover, the continuous functional calculus implies that for every $g \in \mathcal{C}(\mathbb{R})$ and every sequence $U \in \ell^\infty(\mathbb{G};\mathbb{R})$ with $\Vert U \Vert_\infty \leq C$:
\begin{equation} \label{eq_finiterange_norm}
\left\vert \dfrac{1}{\vert \Lambda_L^{(\mathbb{G})}(x) \vert} \mathrm{Tr} \left( P_L^{(\mathbb{G})}(x) g(H_U) P_L^{(\mathbb{G})}(x) \right) \right\vert \leq \Vert g \Vert_{\infty; [-\rho_\mathbb{G} - C , \rho_\mathbb{G} + C]} ~\mbox{.}
\end{equation}
Thus, combining (\ref{eq_finiterange_norm}) - (\ref{eq_finiterangereduction_key}) with the hypothesis in (\ref{eq_finiterange_hypoth}), we estimate
\begin{align}
\left\vert \dfrac{1}{\vert \Lambda_L^{(\mathbb{G})}(x) \vert} \right. & \left. \left[ \mathrm{Tr} \left( P_L^{(\mathbb{G})}(x) f(H_V) P_L^{(\mathbb{G})}(x) \right) - \mathrm{Tr} \left( P_L^{(\mathbb{G})}(x) f(H_{V_W^{(M;x)}})  P_L^{(\mathbb{G})}(x) \right) \right] \right\vert \nonumber \\
   &  \leq \left\vert \dfrac{1}{\vert \Lambda_L^{(\mathbb{G})}(x) \vert} \left[ \mathrm{Tr} \left( P_L^{(\mathbb{G})}(x) \left( f(H_V) - p(H_V) \right) P_L^{(\mathbb{G})}(x) \right) \right. \right\vert 
      \nonumber \\     
    & + \left\vert \left. \dfrac{1}{\vert \Lambda_L^{(\mathbb{G})}(x) \vert} \mathrm{Tr} \left( P_L^{(\mathbb{G})}(x) \left( f(H_{V_W^{(M;x)}}) - p(H_{V_W^{(M;x)}}) \right)  P_L^{(\mathbb{G})}(x) \right) \right] \right\vert \leq 2 \epsilon ~\mbox{,}
\end{align}
thereby establishing the claim.
\end{proof}


\subsection{The Lipschitz property}\label{subsec:LipProp1}

Lemma \ref{lemma_finiterange} reduces changes of the potential to {\em{finite}} subgraphs. Consequently, in order to control the dependence of the  DOSoM on the potential, it suffices, by iteration, to study the variation of the potential at one vertex at a time. To quantify the dependence on the potential at one vertex, while keeping the potential at all other vertices fixed, we will use the following general result about one-parameter families of self-adjoint operators (``\emph{Lipschitz property}''), which we proved in our earlier work \cite[Proposition 3.1]{hislop_marx_2}. To formulate the Lipschitz property, let $T_1, T_2$ be positive, bounded operators and $H_0$ be a (not necessarily bounded) self-adjoint operator on a given Hilbert space $\mathcal{H}$. We consider the one-parameter family of self-adjoint operators
\begin{equation} \label{eq_lipschitz_func_set-up}
H_\lambda := H_0 + \lambda T_1 ~\mbox{, }  \lambda \in \mathbb{R} ~\mbox{.}
\end{equation}
Given a function $f \in \mathcal{C}_c(\mathbb{R})$, we examine the continuity properties of the map
\begin{equation} \label{eq_lipschitz_func}
\mathbb{R} \ni \lambda \mapsto \mathcal{F}_f(\lambda):= \mathrm{Tr}(T_2 f(H_\lambda) T_2) ~\mbox{.}
\end{equation}

To quantify this continuity, we let $\mathrm{Lip}_c(\mathbb{R})$ be the compactly supported, (complex-valued) Lipschitz functions on $\mathbb{R}$ and  let $L_f$ denote the optimal Lipschitz constant of $f \in \mathrm{Lip}_c(\mathbb{R})$, defined by
\begin{equation} \label{eq_lipschitznorm}
L_f:= \sup_{x \neq y \in \R} \dfrac{\vert f(x) - f(y) \vert}{\vert x - y \vert} ~\mbox{.}
\end{equation}
Furthermore, for the map \eqref{eq_lipschitz_func} to be well-defined, we require appropriate trace-ideal conditions for the operators $T_1$ and $T_2$.
In view of this, we let $\mathcal{S}_1(\mathcal{H})$ denote the trace-class operators, $\mathcal{S}_2(\mathcal{H})$ denote the Hilbert-Schmidt operators on $\mathcal{H}$, and write $\Vert . \Vert_j$, $j=1,2$ for the associated Banach space norms.

\begin{prop}[Lipschitz property, Proposition 3.1 in \cite{hislop_marx_2}] \label{prop_lipschitz_traceclass}
Given the setup described in (\ref{eq_lipschitz_func_set-up}) - (\ref{eq_lipschitz_func}), with $T_1, T_2$ positive, bounded operators and $H_0$ a (not necessarily bounded) self-adjoint operator on a given Hilbert space $\mathcal{H}$, with $T_1 \in \mathcal{S}_1(\mathcal{H})$ and $T_2 \in \mathcal{S}_2(\mathcal{H})$. Then for every $f \in \mathrm{Lip}_c(\mathbb{R})$, the map $\lambda \mapsto \mathcal{F}_f(\lambda)$ is Lipschitz in $\lambda$ with
\begin{equation} \label{eq_prop_lipschitz_traceclass}
\left\vert \mathcal{F}_f(\lambda_1) -  \mathcal{F}_f(\lambda_2) \right\vert \leq \min\{ \Vert T_2^2 \Vert ~\Vert T_1 \Vert_{1} ~,~ \Vert T_2 \Vert_{2}^2 ~\Vert T_1 \Vert \} \cdot L_f \cdot \vert \lambda_1 - \lambda_2 \vert ~\mbox{,}
\end{equation}
for each $\lambda_1, \lambda_2 \in \mathbb{R}$, and the constant $L_f$ given in \eqref{eq_lipschitznorm}.
\end{prop}

\begin{remarks}
\begin{itemize}
\item[(i)] While results of this form can be extracted from the literature for test functions $f$ with higher regularity  (cf.\  Alexsandrov and Peller \cite{aleksandrovPeller2011, aleksandrovPeller2016}, and Skripka \cite{skripka2011})), to our knowledge, no explicit proofs for Lipschitz functions $f$ have appeared in the literature. 

\item[(ii)] The main estimate \eqref{eq_prop_lipschitz_traceclass} can be extended to self-adjoint operators $T_1$ that are traceclass but not necessarily positive.  In this case, our proof implies that the upper bound on the right is multiplied by a factor of $3$ (although we do not believe that this factor is necessary). To see this, we write 
$$
T_1 = |T_1| - ( |T_1| - T_1) =: T_1^+ - T_1^- ,
$$
which expresses $T_1$ as the difference of two nonnegative operators $T_1^+$ and $T_1^-$. Following the proof of \cite[Proposition 3.1]{hislop_marx_2}, one sees from \cite[(3.7)]{hislop_marx_2} that this decomposition of $T_1$ leads to two terms and thus two bilinear functionals as in  \cite[(3.9)]{hislop_marx_2}: $\beta^+$ associated with $T_1^+$, and 
$\beta^-$ associated with $T_1^- $, each bilinear functional being defined with respect to a nonnegative operator. Treating each of $\beta^\pm$ separately, as written there, we obtain bounds as in  \cite[(3.15)]{hislop_marx_2} in which $\|T_1 \|_1$ is replaced by $\| T_1^+ \|_1 = \|T_1 \|_1$ for $\beta^+$ and by $\| T_1^- \|_1 \leq 2 \|T_1 \|_1$ for $\beta^-$, and similarly for the operator norms. Combining these terms  yields the factor of $3$ and the result. 

\item[(iii)] The operator $T_2$ need not be positive nor self-adjoint. If $T_2$ isn't self-adjoint, we use $\mathrm{Tr}(T_2^* f(H_\lambda) T_2)$ in \eqref{eq_lipschitz_func}.  

\item[(iv)] In light of remarks (ii)-(iii), we have the following general result that might be of interest in its own right. The proof is similar to the proof of Proposition \ref{prop_lipschitz_traceclass} so we omit it. 
\end{itemize}
\end{remarks}

\begin{prop}[The trace bound] \label{prop_traceclass_general}
Let $T_1, T_2$ be bounded operators and $H_0$ a (not necessarily bounded) self-adjoint operator on a given Hilbert space $\mathcal{H}$. We assume that the self-adjoint, not necessarily positive, operator $T_1 \in \mathcal{S}_1(\mathcal{H})$,
and the  not necessarily self-adjoint operator $T_2 \in \mathcal{S}_2(\mathcal{H})$. We denote by $H$ the self-adjoint operator $H = H_0 + T_1$.  Then for every $f \in \mathrm{Lip}_c(\mathbb{R})$, we have
\begin{equation} \label{eq_traceclass_bound1}
| \mathrm{Tr}(T_2^* f(H) T_2) - \mathrm{Tr}(T_2^* f(H_0) T_2) |  \leq  3  \min \{ \Vert T_2^2 \Vert ~\Vert T_1 \Vert_{1} ~,~ \Vert T_2 \Vert_{2}^2 ~\Vert T_1 \Vert \} \cdot L_f  .
\end{equation}
If $T_1 \geq 0$, the factor ``3'' on the right-hand side of (\ref{eq_traceclass_bound1}) can be replaced by ``1.''
\end{prop}

Returning to the application of Proposition \ref{prop_lipschitz_traceclass}, 
we use the Lipschitz property to quantify changes of the potential at an arbitrary vertex $z \in \mathcal{V}$: For fixed $V \in \ell^\infty(\mathbb{G} \setminus\{z\};\mathbb{R})$, $x \in \mathcal{V}$, and $L >0$, we apply Proposition \ref{prop_lipschitz_traceclass} to the operators,
\begin{equation} \label{eq_applylipschitz}
H_0 = \Delta_\mathbb{G} + \sum_{y \neq z} V(y) \pi_y ~\mbox{, } T_1 = \pi_z ~\mbox{, } T_2 = P_L^{(\mathbb{G})}(x) ~\mbox{,}
\end{equation}
where $\pi_y$ are the rank-one projectors defined in (\ref{eq_elementarycoordproj}). Notice that since $\pi_z$ is a rank-one orthogonal projection, whereas $P_L^{(\mathbb{G})}(x)$ is of rank $\vert \Lambda_L^{(\mathbb{G})}(x) \vert > 1$, we obtain for $T_1$ and $T_2$ as in (\ref{eq_applylipschitz}):
\begin{equation}
\min\{ \Vert T_2^2 \Vert ~\Vert T_1 \Vert_{1} ~,~ \Vert T_2 \Vert_{2}^2 ~\Vert T_1 \Vert \} = \min\{ 1 ~,~ \vert \Lambda_L^{(\mathbb{G})}(x) \vert \} = 1 ~\mbox{.}
\end{equation}
Repeated application of Proposition \ref{prop_lipschitz_traceclass}  for the set-up in \eqref{eq_applylipschitz} and the UGH, specfically (\ref{eq_growthfunction}), thus yield the following:
\begin{lemma} \label{lemma_keyfinitereduction}
Let $x \in \mathcal{V}$ and $M > 0$ be fixed. Then, for all $V,W \in \ell^\infty(\mathbb{G}; \mathbb{R})$ and $L > 0$, one has
\begin{align}
\left\vert \dfrac{1}{\vert \Lambda_L^{(\mathbb{G})}(x) \vert} \right. & \left. \left[ \mathrm{Tr} \left( P_L^{(\mathbb{G})}(x) f(H_W) P_L^{(\mathbb{G})}(x) \right) - \mathrm{Tr} \left( P_L^{(\mathbb{G})}(x) f(H_{V_W^{(M;x)}})  P_L^{(\mathbb{G})}(x) \right) \right] \right\vert \label{eq_lemma_lipschitz_1} \\ 
& \leq L_f \cdot \dfrac{ \vert \Lambda_M^{(\mathbb{G})}(x) \vert}{\vert \Lambda_L^{(\mathbb{G})}(x) \vert} \cdot \Vert V - W \Vert_\infty 
~\mbox{.} \label{eq_lemma_lipschitz_2}
\end{align}
\end{lemma}
We note that by (\ref{eq_modifiedpotential}), the Schr\"odinger operators in the two traces in (\ref{eq_lemma_lipschitz_1}) only differ through the value of the potential within the {\em{finite}} set $\Lambda_M(x)$, therefore giving rise to the numerator of the quotient in (\ref{eq_lemma_lipschitz_2}).


\subsection{Main result on DOSoM}\label{subsec:MainOuter1}

Combining the preparatory results in Lemma \ref{lemma_finiterange} and Lemma \ref{lemma_keyfinitereduction}, we are ready to formulate and prove our main result which establishes continuity of the DOSoM on the underlying potential sequence. To this end, observe that the definition of the DOSoM in (\ref{eq_defnDOSoM}) implies that the map
\begin{equation}
\mathcal{C}(\sigma(H_V)) \ni f \mapsto n_V^*(f) ~\mbox{,}
\end{equation}
defines a sublinear and positively homogeneous map, i.e. a positive sublinear functional on the continuous (complex-valued) functions on $\sigma(H_V)$. More generally, for any finite $M >0$, we will denote the sublinear, complex-valued functionals on $\mathcal{C}([-M,M])$ by $\mathcal{SL}(\mathcal{C}([-M,M]))$, which we equip with the pseudometric $d_w$ defined as follows.
For all $\mu^*, \nu^* \in \mathcal{SL}(\mathcal{C}([-M,M]))$, we define $d_w$ by 
\begin{equation} \label{eq_weakconv_metric}
d_w(\mu^*, \nu^*):= \sup\{ \vert \mu^*(f) - \nu^*(f) \vert ~:~ f \in \mathrm{Lip}([-M,M]) ~\mbox{with } \Vert f \Vert_{\mathrm{Lip}} \leq 1 \} ~\mbox{,}
\end{equation}
where
\begin{equation}
\Vert f \Vert_{\mathrm{Lip}}:= \Vert f \Vert_\infty + L_f ~\mbox{.}
\end{equation}
Our choice of topology is motivated by the well-known fact that weak convergence of measures on $[-M,M]$ is metrizable by the metric given in (\ref{eq_weakconv_metric}), which is also known as the Fortet-Mourier metric; see e.g. \cite{Dudley_1966, Dudley_1976_book}, and \cite[Chapter 6]{CVillani_optimaltransport_book} for a more recent account. In particular, for models for which the DOSoM is actually a measure, i.e. for which a density of states measure exists, the topology induced by the pseudometric in (\ref{eq_weakconv_metric}) will ensure that our result implies continuity of the DOSm in the underlying potential sequence (and $\ell^\infty$ norm) with respect to weak topology of measures.

\begin{theorem} \label{thm_DOSoM_main}
Let $\mathbb{G}=(\mathcal{V},\mathcal{E})$ be an infinite, connected graph satisfying the UGH. Then, for each finite $C>0$, the map
\begin{equation} \label{eq_mainthmmap}
\mathcal{N}_C: \ell^\infty(\mathbb{G};[-C,C]) \to \mathcal{SL}(\mathcal{C}([-\rho_\mathbb{G} - C, \rho_\mathbb{G} + C])) ~\mbox{, } V \mapsto n_V^*
\end{equation}
is continuous with respect to the topology for the codomain induced by (\ref{eq_weakconv_metric}). The modulus of continuity is quantified by the following: if $\gamma_\mathbb{G}$ is a uniform growth function for $\mathbb{G}$
in the sense of (\ref{eq_growthfunction}), then for every $\zeta >0$ one has 
\begin{equation} \label{eq_mainthm1}
d_w(n_V^*, n_W^*) \leq 2^{3/2} (\rho_\mathbb{G} + C) c_b \cdot \dfrac{1}{\sqrt{  \gamma_\mathbb{G}^{-1}\left(\left(\dfrac{1}{\Vert V - W \Vert_\infty}\right)^\zeta \right) }    } +  \left( \Vert V - W \Vert_\infty \right)^{1 - \zeta}
\end{equation}
for all $V, W \in \ell^\infty(\mathbb{G};[-C,C])$ with $\Vert V - W \Vert_\infty < 1$; here, $c_b$ is an absolute constant given in (\ref{eq:bernsteinconst}). An analogous result holds for $n_{V;x}^*$, the local DOSoM \eqref{eq:localDOSoM4}, for all $x \in \mathcal{V}$.
\end{theorem}



As in our earlier work \cite{hislop_marx_1}, the proof of Theorem \ref{thm_DOSoM_main} will use polynomial approximation of Lipschitz functions $f \in  \mathrm{Lip}([-\rho_\mathbb{G} - C, \rho_\mathbb{G} + C])$ by Bernstein polynomials. In view of this, given $g \in \mathcal{C}([0,1])$, we denote by $B_n[g](x)$ the $n^{th}$ Bernstein polynomial associated with $g$,
\begin{equation}
B_n[g](x) = \sum_{k=0}^n  \binom{n}{k} g \left( \frac{k}{n} \right) x^k (1-x)^{n-k} ~\mbox{.}
\end{equation}
It is well-known (see e.g. \cite{beals1}) that for $g \in C([0,1])$ with modulus of continuity $W_g$ on $[0,1]$, the approximation by $B_n[g]$ satisfies
\beq\label{eq:poly-approx1}
\| B_n[g] - g \|_\infty \leq c_b W_g(n^{-1/2}) ~\mbox{,}
\eeq
where $c_b$ is an absolute constant. Here, as usual, a modulus of continuity $W_g : [0,1] \rightarrow [0, \infty)$, for $g \in C([0,1])$, is defined  by $|g(x) - g(y)| \leq W_g(|x-y|)$, for $x , y 
\in [0,1]$. It is well known that the $n^{-1/2}$ dependence in (\ref{eq:poly-approx1}) is, in general, optimal \cite{sikkema} in which case the optimal value for the constant in (\ref{eq:poly-approx1}) is given by
\begin{equation} \label{eq:bernsteinconst}
c_b = \dfrac{4306 + 837 \sqrt{6}}{5832} \approx 1.08989 ~\mbox{.} 
\end{equation}

\begin{proof}
Let $V \neq W \in \ell^\infty(\mathbb{G};[-C,C])$ with $0 < \Vert V - W \Vert_\infty =:\epsilon < 1$ be given and $f \in \mathrm{Lip}([-\rho_\mathbb{G} - C, \rho_\mathbb{G} + C])$ be fixed and arbitrary. Further, let $\eta: (0, +\infty) \to (0, + \epsilon)$ be a function, determined appropriately later, which satisfies $\eta(y) \searrow 0$ as $y \searrow 0^+$.

Given $f \in \mathrm{Lip}([-\rho_\mathbb{G} - C, \rho_\mathbb{G} + C])$, we let $p: [-\rho_\mathbb{G} - C, \rho_\mathbb{G} + C] \to \mathbb{C}$ be a Bernstein polynomial with domain rescaled from $[0,1]$ to $[-\rho_\mathbb{G} - C, \rho_\mathbb{G} + C]$ and with degree $\deg(p) =: n$, chosen to ensure that 
\begin{equation} \label{eq_maintheorem_approx}
\Vert f - p \Vert_\infty <\frac{1}{2} \eta(\epsilon) ~\mbox{.}
\end{equation}
Taking into account the rescaling of the domain of the Bernstein polynomial, the rate of convergence in (\ref{eq:poly-approx1}) implies that (\ref{eq_maintheorem_approx}) is satisfied if we take 
\begin{equation} \label{eq_mainthm_degree}
n = \left\lceil \left(\dfrac{4 (\rho_\mathbb{G} + C) c_b L_f}{\eta(\epsilon)} \right)^2 \right\rceil ~\mbox{.}
\end{equation}

Thus, for all $x \in \mathcal{V}$ and $L \in \mathbb{N}$, the combination of Lemma \ref{lemma_finiterange} with $M = L + \lfloor n/2 \rfloor$ as therein and Lemma \ref{lemma_keyfinitereduction} yields
\begin{align} \label{eq_mainthm_estimkey}
\left\vert \dfrac{1}{\vert \Lambda_L^{(\mathbb{G})}(x) \vert} \right. & \left. \left[ \mathrm{Tr} \left( P_L^{(\mathbb{G})}(x) f(H_V) P_L^{(\mathbb{G})}(x) \right) - \mathrm{Tr} \left( P_L^{(\mathbb{G})}(x) f(H_{W})  P_L^{(\mathbb{G})}(x) \right) \right] \right\vert \nonumber \\
\leq & ~\left\vert \dfrac{1}{\vert \Lambda_L^{(\mathbb{G})}(x) \vert} \left[ \mathrm{Tr} \left( P_L^{(\mathbb{G})}(x) f(H_V) P_L^{(\mathbb{G})}(x) \right) - \mathrm{Tr} \left( P_L^{(\mathbb{G})}(x) f(H_{V_W^{(M;x)}})  P_L^{(\mathbb{G})}(x) \right) \right] \right\vert + \nonumber \\
& \left\vert \dfrac{1}{\vert \Lambda_L^{(\mathbb{G})}(x) \vert} \left[ \mathrm{Tr} \left( P_L^{(\mathbb{G})}(x) f(H_{V_W^{(M;x)}}) P_L^{(\mathbb{G})}(x) \right) - \mathrm{Tr} \left( P_L^{(\mathbb{G})}(x) f(H_{W})  P_L^{(\mathbb{G})}(x) \right) \right] \right\vert \nonumber \\
\leq & ~\eta(\epsilon) + L_f \cdot \dfrac{ \vert \Lambda_M^{(\mathbb{G})}(x) \vert}{\vert \Lambda_L^{(\mathbb{G})}(x) \vert} \cdot \epsilon ~\mbox{.} 
\end{align}
Taking the supremum over all $x \in \mathcal{V}$ and, subsequently, the $\limsup$ as $L \to \infty$, we thus conclude
\begin{equation}
\vert n_V^*(f) - n_W^*(f) \vert \leq ~\eta(\epsilon) + L_f \cdot \gamma_\mathbb{G}\left( \left\lfloor \frac{n}{2} \right\rfloor \right) \cdot \epsilon ~\mbox{.} \label{eq_mainthm_estim1}
\end{equation}

In particular, letting
\begin{equation} \label{eq_thmmaineta}
\eta(y) = 2^{3/2} (\rho_\mathbb{G} + C) c_b \cdot L_f  \cdot \sqrt{ \dfrac{1}{\gamma_\mathbb{G}^{-1}(\frac{1}{y^\zeta} )  }} ~\mbox{, for $\zeta>0$ ,}
\end{equation}
(\ref{eq_mainthm_estim1}) yields the rightmost side of (\ref{eq_mainthm1}), thereby verifying the claim.
\end{proof}


\section{Applications to  deterministic Schr\"odinger operators on $\mathbb{Z}^d$ and the Bethe lattice}\label{sec:applications1}
\setcounter{equation}{0}

In this section, we present several applications of Theorem \ref{thm_DOSoM_main} to two particularly interesting graphs: 1) the $d$-dimensional lattice $\mathbb{G} = \mathbb{Z}^d$,  and 2)  the Bethe lattice $\mathbb{G} = {\mathbb{B}_k}$ with coordination number $k \geq 3$. We also include consequences for the integrated outer density of states (IoDS, defined in \eqref{def_IDS}) and the weak coupling limit of both of these examples. 



\subsection{Deterministic Schr\"odinger operators on $\mathbb{Z}^d$}

We recall that by (\ref{eq_latticegrowth}), for the $d$-dimensional lattice $\mathbb{G} = \mathbb{Z}^d$, {\em{any}} strictly increasing function $\gamma_\mathbb{G}: [1, + \infty) \to [1, +\infty)$ can serve as a uniform local growth function in the sense of (\ref{eq_growthfunction}). Given this flexibility, we apply Theorem \ref{thm_DOSoM_main} with the choice of the uniform growth function given by
\begin{equation} \label{eq_growthlattice}
\gamma_{\mathbb{Z}^d}(y) = y^{\zeta/\alpha},
\end{equation}
for parameters $\zeta, \alpha > 0$ positive. As a consequence, we obtain that for all $V \neq W \in \ell^\infty(\mathbb{\Z}^d;[-C,C])$, with $0 < \epsilon:= \Vert V - W \Vert_\infty < 1$, the following bound
\begin{equation}\label{eq:DOSoM5}
d_w(n_V^*, n_W^*) \leq 2^{3/2} (2d + C) c_b \cdot \epsilon^{\zeta/\alpha} + \epsilon^{1 - \zeta} ~\mbox{.}
\end{equation} 
In particular, letting $\alpha \to 0^+$ in \eqref{eq:DOSoM5}, yields the bound
\begin{equation}
d_w(n_V^*, n_W^*) \leq \epsilon^{1 - \zeta} ~\mbox{.}
\end{equation} 
Thus, taking $\zeta \to 0^+$, we arrive at:

\begin{theorem} \label{thm_lattice_DOSoM}
Consider the $d$-dimensional lattice $\mathbb{G} = \mathbb{Z}^d$, $d \in \mathbb{N}$. Then, for each fixed $C>0$, the map in (\ref{eq_mainthmmap}) is Lipschitz continuous with respect to the topology for the codomain induced by the pseudometric in (\ref{eq_weakconv_metric}), i.e. for all $V, W \in \ell^\infty(\mathbb{Z}^d;[-C,C])$ with $\Vert V - W \Vert_\infty < 1$, one has
\begin{equation} \label{eq_mainthm_zd}
d_w(n_V^*, n_W^*) \leq \Vert V - W \Vert_\infty ~\mbox{.}
\end{equation}
\end{theorem}

We remark that this result on the DOSoM holds for any lattice, such as a hexagonal or triangular lattice, for which the uniform growth function may be chosen as in \eqref{eq_growthlattice}.

We present two applications of Theorem \ref{thm_lattice_DOSoM}: 1) the continuity of the cumulative distribution function
associated with the DOSoM for a potential sequence, which we call the\emph{ integrated outer density of states} (IoDS), and 2) upper bounds on the  DOSoM and IoDS as functions of the disorder in the weak coupling limit for lattice Schr\"odinger operators on $\mathbb{Z}^d$.


\subsubsection{Consequences for the integrated outer density of states for lattice Schr\"odinger operators}\label{subsec:DOSoM1}

Given $V \in \ell^\infty(\mathbb{Z}^d; \mathbb{R})$ we define the {\em{integrated outer density of states}} (IoDS) as the cumulative distribution associated with DOSoM $n_V^*$, i.e.
\begin{equation} \label{def_IDS}
N_V^*(E):= n_V^*(\chi_{(-\infty ,E]}) ~\mbox{.}
\end{equation}
We recall the known continuity properties of the DOSoM \emph{ with respect to the energy } for {\em{fixed potential}}, which were established for lattice Schr\"odinger operators in \cite[Theorem 2.2]{bourgain-klein1}. Bourgain and Klein
proved that for each fixed $V \in \ell^\infty(\mathbb{Z}^d; \mathbb{R})$ with $\Vert V \Vert_\infty \leq C$, there exits a constant $K_{d; C}> 0$, depending only on $d$ and $C$, such that the DOSoM is $\log$-H\"older continuous, i.e.\ for each $E \in \mathbb{R}$ and $0 < \epsilon \leq \frac{1}{2}$, one has
\begin{equation} \label{eq_BK_loghold}
n_V^*([E, E+ \epsilon]) \leq \dfrac{K_{d; C}}{\log \left( \frac{1}{\epsilon} \right)} ~\mbox{.}
\end{equation}
For the special case of Schr\"odinger operators on $\mathbb{Z}^d$, we recall that (\ref{eq_compareBKdefn}) implies that our definition of the DOSoM in (\ref{eq_defnDOSoM}) 
reduces to the definition given in \cite{bourgain-klein1}.

The results for the DOSoM in Theorem \ref{thm_DOSoM_main} and the known continuity properties in the energy given in (\ref{eq_BK_loghold}) imply the following bound for the IoDS:

\begin{theorem} \label{thm_IDS_lattice}
Consider the $d$-dimensional lattice $\mathbb{G} = \mathbb{Z}^d$, $d \in \mathbb{N}$.  For each fixed $C>0$, there exists a constant $K_0 = K_0(d, C)$ such that for all $V, W \in \ell^\infty(\mathcal{Z}^d;[-C,C])$ with $\Vert V - W \Vert_\infty < 1$, one has
\begin{equation} \label{eq_lattice_ids}
\vert N_V^*(E) - N_W^*(E) \vert \leq \dfrac{K_0}{\log \left( \frac{1}{\Vert V - W \Vert_\infty} \right)} ~\mbox{, for all $E \in \mathbb{R}$.}
\end{equation}
\end{theorem}

\begin{remarks}\label{remark:iods1}
\begin{itemize}
\item[(i)] The constant $K_0$ is given explicitly by the numerator of the right hand side of (\ref{eq_ids_mainestim_final}).
\item[(ii)] As can be seen from the proof below, the modulus of continuity in (\ref{eq_lattice_ids}) is limited by the modulus of continuity of the IoDS in the energy, the latter of which is in general known to only be log-H\"older continuous as given in (\ref{eq_BK_loghold}). In particular, the proof below shows that, if for a potential $V$ the IoDS is known to be $\alpha$-H\"older continuous, $0 < \alpha < 1$, at an anergy $E \in \mathbb{R}$ under consideration, then the $\alpha$-H\"older continuity will be inherited by the modulus of continuity for the map $W \mapsto N_W^*(E)$ for potentials $W$ in a neighborhood of $V$. This will play a role for the weak-coupling behavior discussed in section \ref{subsec_lattice_weak}.
\end{itemize}
\end{remarks}

\begin{proof}
Let $V, W \in \ell^\infty(\mathbb{Z}^d; [-C,C])$ with 
\begin{equation}
0 < \epsilon:= \Vert V - W \Vert_\infty < \frac{1}{2} ~\mbox{, } 
\end{equation}
be given. Since $N_V^*(E) = N_W^*(E)$ for $\vert E \vert \geq C + 2d$, it suffices to consider $E \in (-2d - C, 2d + C)$. 
Fix $\zeta > 0$ to be determined appropriately later. Following a similar approach as in our proof of \cite[Theorem 3.2]{hislop_marx_1}, we approximate the characteristic function $\chi_{(-\infty, E]}$ on $[-2d - C, 2d + C]$ by the Lipschitz functions $f_{\epsilon; \zeta}^{(\pm)}: [-2d - C, 2d + C] \to \mathbb{R}$, defined, respectively, by
\beq\label{eq:cut-off1}
f_{\epsilon; \zeta}^{(-)} (x) := \left\{ \begin{array}{ll}
                     1 & {\rm ,  if} ~~ x \in [- 2d - C, E- \epsilon^\zeta/2] ~\mbox{, } \\
                      1 - \frac{1}{\epsilon^\zeta/2} (x- (E-\epsilon^\zeta/2)) &  {\rm , if} ~~x \in (E-\epsilon^\zeta/2, E] ~\mbox{, }  \\
                      0 & {\rm , if} ~~ x > E ~\mbox{, } 
                      \end{array}
                      \right.
 \eeq
and
\beq\label{eq:cut-off2}
f_{\epsilon; \zeta}^{(+)} (x) := \left\{ \begin{array}{ll}
                     1 & {\rm , if} ~~ x \in [- 2d - C, E] ~\mbox{, }  \\
                      1 - \frac{1}{\epsilon^\zeta/2} (x- E) &  {\rm , if} ~~x  \in (E, E+\epsilon^\zeta/2 ] ~\mbox{, }  \\
                      0 & {\rm , if} ~~ x > E + \epsilon^\zeta/2 ~\mbox{. } 
                      \end{array}
                      \right.
 \eeq
 
For $x \in \mathbb{Z}^d$ and $L \in \mathbb{N}$, we consider the local DOSoM
\begin{equation}
n_{V;L}^{(x)}(f) := \dfrac{1}{\vert \Lambda_L^{(\mathbb{Z}^d)}(x) \vert} \mathrm{Tr}\left( P_L^{(\mathbb{Z}^d)}(x) f(H_V) P_L^{(\mathbb{Z}^d)}(x) \right), 
\end{equation}
for any bounded Borel function $f$. 
Note that, by construction, one has
\begin{equation} \label{eq_approx_IDS}
0 \leq f_{\epsilon; \zeta}^{(-)} \leq \chi_{[- C - 2d, E)} \leq f_{\epsilon; \zeta}^{(+)} ~\mbox{, }  L_{f_{\epsilon; \zeta}^{(\pm)}} = \frac{2}{\epsilon^\zeta} ~\mbox{.}
\end{equation}
Thus, in particular, for each $x \in \mathbb{Z}^d$ and $L \in \mathbb{N}$, (\ref{eq_approx_IDS}) implies
\begin{equation} \label{eq_IDS_localmeas_approx}
\vert n_{V;x}^{(L)}(\chi_{(-\infty ,E]}) - n_{W;x}^{(L)}(\chi_{(-\infty ,E]})   \vert \leq \max_{\pm} | n_{V;x}^{(L)}(f_{\epsilon; \zeta}^{(\pm)}) -  n_{W;x}^{(L)}(f_{\epsilon; \zeta}^{(\pm)}) |  + n_{W;x}^{(L)}(\chi_{[E-\epsilon^\zeta/2, E+\epsilon^\zeta/2]}) ~\mbox{.}
\end{equation}

For $\alpha >0$, $\gamma_{\mathbb{Z}^d}$ as in (\ref{eq_growthlattice}), and $f = f_{\epsilon; \zeta}^{(\pm)}$, we take $n_{\epsilon; \zeta}^{(\pm)} \in \mathbb{N}$ as in (\ref{eq_mainthm_degree}) and $\eta_{\epsilon; \zeta}^{(\pm)}$ as in (\ref{eq_thmmaineta}). Letting 
\begin{equation}
M_{\epsilon; \zeta}^{(\pm)}:= L + \lfloor n_{\epsilon; \zeta}^{(\pm)}/2 \rfloor ~\mbox{,}
\end{equation}
application of the analogue of (\ref{eq_mainthm_estimkey}) to the first term on the right hand side of (\ref{eq_IDS_localmeas_approx}), yields
\begin{align}\label{eq:iods1}
\vert n_{V;x}^{(L)}(\chi_{(-\infty ,E]}) - n_{W;x}^{(L)}(\chi_{(-\infty ,E]}) \vert \leq & ~2^{3/2} (2d + C) c_b \cdot L_{f_{\epsilon; \zeta}^{(\pm)}} \cdot \epsilon^{\zeta/\alpha} +  L_{f_{\epsilon; \zeta}^{(\pm)}} \cdot \dfrac{ \vert \Lambda_{M_{\epsilon; \zeta}^{(\pm)}}^{(\mathbb{Z}^d)}(x) \vert}{\vert \Lambda_L^{(\mathbb{Z}^d)}(x) \vert} \cdot \epsilon \nonumber \\
& + n_{W;x}^{(L)}(\chi_{[E-\epsilon^\zeta/2, E+\epsilon^\zeta/2]}) ~\mbox{.}
\end{align}
In particular, by (\ref{eq_approx_IDS}) and (\ref{eq_BK_loghold}), taking the supremum over $x \in \mathbb{Z}^d$ followed by the $\limsup$ as $L \to \infty$ in \eqref{eq:iods1}, we arrive at
\begin{align} \label{eq_ids_mainestim_1}
\vert N_V^*(E) - N_W^*(E) \vert \leq & ~\frac{2}{\epsilon^\zeta} \left( 2^{3/2} (2d + C) c_b \cdot \epsilon^{\zeta/\alpha} +  \epsilon^{1 - \zeta} \right) + n_W^*(\chi_{[E-\epsilon^\zeta/2, E+\epsilon^\zeta/2]}) \nonumber \\
\leq & ~\frac{2}{\epsilon^\zeta} \left( 2^{3/2} (2d + C) c_b \cdot \epsilon^{\zeta/\alpha} +  \epsilon^{1 - \zeta} \right) + \frac{K_{d; C}}{\log(\epsilon^{-\zeta})}  ~\mbox{.}
\end{align}
Thus, letting $\alpha \to 0^+$, (\ref{eq_ids_mainestim_1}) produces
\begin{equation} \label{eq_ids_mainestim_2}
\vert N_V^*(E) - N_W^*(E) \vert \leq ~2 \epsilon^{1 - 2 \zeta} + \frac{K_{d; C}}{\log(\epsilon^{-\zeta})} ~\mbox{,}
\end{equation}
Finally, since $y^\beta \geq \log y$ for all $y \in (0, + \infty)$ if and only if $\beta \geq \frac{1}{\mathrm{e}}$, we optimize $\zeta > 0$, giving $\zeta = (2 + \mathrm{e})^{-1}$, and consequently obtain
\begin{equation} \label{eq_ids_mainestim_final}
\vert N_V^*(E) - N_W^*(E) \vert \leq \dfrac{2 (2 + \mathrm{e}) \max\{2; K_{d;C} \}}{\log \left( \frac{1}{\epsilon} \right) } ~\mbox{.}
\end{equation}
\end{proof}


\subsubsection{Consequences for the weak coupling limit for lattice Schr\"odinger operators} \label{subsec_lattice_weak}

The weak disorder limit of the DOSm and IDS for random lattice Schr\"odinger operators was studied in \cite[section 5.1]{hislop_marx_1}. 
There, we considered random lattice Schr\"odinger operators of the form $H_\omega (\lambda) = \Delta_{\Z^d} + \lambda \sum_{x \in \Z^d} \omega_x \pi_x$, where the sequence $\omega \in \Omega = [-1,1]^{\ZZ^d}$ consists of $iid$ random variables with common, arbitrary probability measure $\mu$ supported on $[-1,1]$. By rescaling the random variables $\widetilde{\omega}_x := \lambda \omega_x$, for $x \in \Z^d$,  we applied the strategy of \cite{hislop_marx_1} to compare the Hamiltonian $H_{\widetilde{\omega}}$, with the rescaled single-site probability measure $\mu_\lambda$, with $H_0 := \Delta$, as $\lambda \rightarrow 0^+$.  
Our two main results were
\begin{enumerate}
\item DOSm: For all Lipschitz functions $f$, 
$$
| n_\lambda^{(\infty)}(f) - n_{\lambda = 0}^{(\infty)} (f) |  \leq \gamma \| f \|_{\rm Lip} \lambda^{\frac{1}{1 + 2d}},
$$
\item IDS:
\beq\label{eq:latticeIDS1}
| N_\lambda (E) - N_{\lambda = 0}(E) | \leq c \lambda^{ \left( \frac{1}{1 + 2d} \right) \left( \frac{\delta}{1 + \delta} \right) }, 
\eeq
where $\delta = \frac{1}{2}$ for $d=1$, and $\delta =1$, for $d\geq 2$.
\end{enumerate}

We now apply this strategy to the deterministic case presented in section \ref{subsec:DOSoM1}.
Due to the improvement in Theorem \ref{thm_lattice_DOSoM}, we find that by replacing $V \in \ell^\infty ( \Z^d; [-C, C])$ 
by $\lambda V$, and taking $W =0$, we obtain Lipschitz continuity with respect to $\lambda$ of the DOSoM.

\begin{corollary}\label{cor:contDOSoMweak1}
For each potential sequence $V \in \ell^\infty ( \ZZ^d; [-C, C])$, the DOSoM is continuous with respect to the coupling parameter $\lambda > 0$ and satisfies the bound
\beq\label{eq:dosmWeak1}
d_w (n_{\lambda V}^*, n_0 ) \leq \lambda \|V \|_\infty .
\eeq
An analogous result holds for the local DOSoM $n_{\lambda V;x}^*$, for all $x \in \mathbb{Z}^d$.
\end{corollary}

Corollary \ref{cor:contDOSoMweak1} implies an improvement on the modulus of continuity of the IoDS associated with the DOSoM. As in the proof of the weak disorder continuity of the IDS for random Schr\"odinger operators in \cite[Theorem 5.2]{hislop_marx_1}, the bound \eqref{eq:latticeIDS1} on the IDS depends on the H\"older continuity of the IDS in energy for the lattice Laplacian. As in \cite[(5.9)]{hislop_marx_1}, 
\beq\label{eq:IDSfree1}
| N_{\lambda = 0}(E + \epsilon) - N_{\lambda = 0}(E) | \leq c_0 \epsilon^\delta, 
\eeq
 where, for $d=1$, $\delta = \frac{1}{2}$, and for $d \geq 2$, we may take $\delta = 1$. Combining \eqref{eq:dosmWeak1} with \eqref{eq:IDSfree1}, and recalling (ii) of Remarks \ref{remark:iods1},  we find that for the general setting:

\begin{corollary}\label{cor:IoDSweak1}
For each potential sequence $V \in \ell^\infty ( \ZZ^d; [-C, C])$, the IoDS is continuous with respect to the coupling parameter $\lambda > 0$ and satisfies the bound
\beq\label{eq:IDSoF1}
| N_{\lambda V}^* (E) - N_0(E) | \leq c \lambda^{ \left( \frac{\delta}{1+\delta} \right)},
\eeq
so for $d =1$, the modulus of continuity is $\lambda^{\frac{1}{3}}$, and for $d \geq 2$, it is $\lambda^{\frac{1}{2}}$.
\end{corollary}

As discussed in section \ref{subsec:RSO1}, Corollary \ref{cor:IoDSweak1} can be applied to random Schr\"odinger operators
and yields an improvement over our  results in \cite{hislop_marx_1}. Prior to \cite{hislop_marx_1}, Schenker \cite{schenker} and Hislop, Klopp, and Schenker \cite{hks} proved bounds on the modulus of continuity for the IDS for random lattice Schr\"odinger operators in any dimension under the assumption that the single-site probability measure $\mu$ has a bounded density with compact support. 
The bound in \eqref{eq:IDSoF1} improves the $\lambda^{\frac{1}{8}}$-dependence attained in earlier works \cite{hks,schenker} and removes any assumptions on the single-site probability measure.



\subsection{Deterministic Schr\"odinger operators on the Bethe lattice}

We recall that for the Bethe lattice with coordination number $k \geq 3$, the spectral radius for the Laplacian is 
\begin{equation}
\rho_{\mathbb{B}_k} = 2 \sqrt{k-1} ~\mbox{.}
\end{equation}
Thus, Theorem \ref{thm_DOSoM_main} immediately yields the following:

\begin{theorem} \label{thm_Bethe}
Consider the Bethe lattice $\mathbb{G} = \mathbb{B}_k$ with coordination number $k \geq 3$. Then, for each $C > 0$, the map in \eqref{eq_mainthmmap} is $\frac{1}{2}$-log-H\"older continuous with respect to the topology for the codomain induced by the pseudometric in (\ref{eq_weakconv_metric}), i.e.\ there exists a constant $\gamma_k > 0$, explicitly given in (\ref{eq_bethe_constant}), such that for all $V, W \in \ell^\infty(\mathbb{B}_k;[-C,C])$ with $\Vert V - W \Vert_\infty < 1$, one has
\begin{equation} \label{eq_thm_Bethe_modulus}
d_w(n_V^*, n_W^*) \leq \dfrac{\gamma_k}{\sqrt{\log\left(\frac{1}{ \Vert V - W \Vert_\infty}    \right)     }} ~\mbox{.}
\end{equation}
\end{theorem}

We note that the modulus of continuity in Theorem \ref{thm_Bethe} reproduces our earlier result for random Schr\"odinger operators on $\mathbb{B}_k$ from \cite[Theorem 6.1]{hislop_marx_1}.

\begin{proof}
As earlier, we write $0< \epsilon: = \Vert V - W \Vert_\infty < 1$. We use the optimal local growth function for $\mathbb{B}_k$ given in (\ref{eq_bethegrowth}), i.e. $\gamma_{\mathbb{B}_k}(n) = (k-1)^n$. Then, for each $\zeta > 0$, using that \begin{equation} \label{eq_bethe_constant}
\frac{ \gamma_k}{2}:= \sqrt{\log(k-1)} 2^{3/2} ( \rho_{\mathbb{B}_k} + C) c_b > 8 \sqrt{\log(2)} > 1 ~\mbox{, }
\end{equation}
for every $C > 0$ and $k \geq 3$, (\ref{eq_mainthm1}) of Theorem  \ref{thm_DOSoM_main} yields
\begin{align}
d_w(n_V^*, n_W^*) & \leq \frac{ \gamma_k}{2} \left( \dfrac{1}{\sqrt{ \log(\epsilon^{-\zeta}})    } + \epsilon^{1 - \zeta}  \right)  \nonumber \\
   & = \frac{1}{2} \gamma_k \left( \dfrac{1}{\sqrt{ \log(\epsilon^{-\zeta} )   } } + \dfrac{1}{\sqrt{(\epsilon^{- \zeta})^{2 (1 - \zeta)/\zeta}}}  \right) ~\mbox{.}
\end{align}
Finally, as in the proof of Theorem \ref{thm_IDS_lattice}, we use that $y^\beta \geq \log y$ for all $y \in (0, + \infty)$ if and only if $\beta \geq \frac{1}{\mathrm{e}}$, which determines the optimal value of $\zeta$ as
\begin{equation}
\zeta = \dfrac{2 e}{1 + 2 e} ~\mbox{.}
\end{equation}
With this choice, we obtain the upper bound  (\ref{eq_thm_Bethe_modulus}).
\end{proof}

\section{Applications to  ergodic and random Schr\"odinger operators on graphs}\label{sec:applications2}
\setcounter{equation}{0}

The methods developed here for deterministic Schr\"odinger operators operators have immediate consequences for general families of ergodic Schr\"odinger operators on graphs. A subset of these operators are random Schr\"odinger operators. Another implication of Theorem \ref{thm_lattice_DOSoM} for random Schr\"odinger operators on the lattice is Lipschitz continuity of the DOSm in the underlying single-site probability measure. This is an improvement of \cite[Theorem 3.1]{hislop_marx_1}, where H\"older continuity was proven, and recovers the results of Shamis \cite{shamis} and of Kachkovskiy \cite{kachkovskiy}. 

\subsection{Ergodic Schr\"odinger operators}\label{subsec:ergodicSO1}

In this section, we apply our results to ergodic Schr\"odinger operators (defined after \eqref{eq:covariance1}) on lattices $\Z^d$ and on the Bethe lattice $\mathbb{B}_k$.  We begin with a general formulation on a graph $\mathbb{G} = ( \mathcal{V}, \mathcal{E})$.
 Let $(\Omega, \mathcal{B}, \Pp)$ be a probability space with a family $\mathcal{T} :=  \{ T_x ~|~ x \in \mathcal{V} \}$ of invertible maps $T_x: \Omega \rightarrow \Omega$. We recall that the family $\mathcal{T}$ of invertible maps $T_x$ is called \emph{ergodic} if, for all $x \in \mathcal{V}$, the measure $\mu$ is $T_x$-invariant, and if all $\mathcal{T}$-invariant sets in $\mathcal{B}$ satisfy a $0-1$ law.


 We also assume that there is a family of maps
 $\{ \alpha_x : \mathbb{G} \rightarrow \mathbb{G}, l x \in \mathcal{V} \}$, which is  a subgroup of the automorphism group ${\rm Aut}(\mathbb{G})$. We further assume that the transformations $T_x$ satisfy the following covariance relation with respect to the maps $\alpha_x$:
\beq\label{eq:automCovar1}
T_x \circ T_{x'} = T_{\alpha_{x'}(x)}, ~~~~\forall x, x' \in \mathcal{V} , 
\eeq
and that the family of maps  $\{ \alpha_x \}$ satisfy an associated consistency relation: 
\beq\label{eq:automCovar2}
\alpha_{x'} \circ \alpha_x = \alpha_{\alpha_{x'}(x)} .
\eeq

We consider the Hilbert space $\mathcal{H} := \ell^2 (\mathbb{G}; \mathbb{C})$. We can construct a unitary representation of the family $\mathcal{T}$ using the maps $\alpha_x$ as follows. For any $\psi \in \mathcal{H}$, we define
\beq\label{eq:Taction1}
(U_{T_x} \psi)(x') := \psi (\alpha_{x}(x')),  ~~{\rm and} ~~ (U_{T_x^{-1}} \psi )(x') :=  \psi (\alpha_{x}^{-1}(x')) .
\eeq
 Due to the consistency relations \eqref{eq:automCovar1} and \eqref{eq:automCovar2}, this family $\{ U_{T_x} ~|~ x \in \mathcal{V} \}$  is a unitary representation of the family $\mathcal{T}$ of invertible ergodic maps $T_x$. In particular, one checks that 
$$
U_{T_x}^{-1} = U_{T_x^{-1}} = U_{T_x}^* .
$$  

We suppose that we have a weakly $\Pp$-measurable map $\omega \in \Omega \rightarrow \{ H_\omega \}$ into the bounded linear operators on $\mathcal{H}$. We say that this map is covariant with respect the ergodic family $\mathcal{T} = \{ T_x ~|~ x \in \mathcal{V} \}$ if
\beq\label{eq:covariance1}
U_{T_x} H_\omega U_{T_x}^* = H_{T_x (\omega)},
\eeq
for all $x \in \mathcal{V}$. 
An \emph{ergodic Schr\"odinger operator}  (ESO) is a weakly $\Pp$-measurable map $\omega \in \Omega \rightarrow \{ H_\omega \} \subset \mathcal{B} ( \ell^2 (\mathbb{G}; \mathbb{C} ))$ for which the bounded operators $H_\omega$ satisfying the covariance relation \eqref{eq:covariance1}.

The basic construction of a covariant family of discrete generalized  Schr\"odinger operators $H_\omega$ on $\mathcal{H}$ is as follows. 
For a function $v \in L^\infty ( (\Omega, \Pp); \R)$, called a\emph{ sampling function}, we define a potential
\beq\label{eq:ErgodicPot1}
V_\omega (x) := v(T_x (\omega)), ~~ {\rm for} ~~ \omega \in \Omega .
\eeq
It is easy to check that definition \eqref{eq:Taction1} and conditions \eqref{eq:automCovar1} and \eqref{eq:automCovar2} guarantee the covariance relation
\beq\label{eq:potentialCovar1}
U_{T_x} V_\omega U_{T_x}^* = V_{T_x(\omega)}
\eeq
Let $L_\mathbb{G}$ be a self-adjoint,  finite-difference operator on the graph $\mathbb{G}$, like the Laplacian $\Delta_\mathbb{G}$. We require that $L_\mathbb{G}$ be $\mathcal{T}$-invariant in that
\beq\label{eq:LaplaceInv1}
U_{T_x} L_\mathbb{G} U_{T_x}^* = L_\mathbb{G}, ~~~ \forall x \in \mathcal{V}.
\eeq
Then, from \eqref{eq:potentialCovar1} and \eqref{eq:LaplaceInv1}, the (bounded) self-adjoint, Schr\"odinger operator 
\begin{equation} \label{eq_ESO_1}
H_\omega := \Delta_\mathbb{G} + V_\omega
\end{equation}
satisfies the covariance relation \eqref{eq:covariance1} and is an  ESO. For simplicity, we will will take $L_{\mathbb{G}} = \Delta_\mathbb{G}$ in what follows.  Examples of ESO include almost-periodic and limit-periodic Schr\"odinger operators, quasi-periodic Schr\"odinger operators, and random Schr\"odinger operators. For an introduction to ergodic Schr\"odinger operators on $\Z^d$, we refer the reader to \cite[Chapter 3]{aizenmanWarzel_book} and \cite[Chapter 9]{cfks}.  

The following theorem is a consequence of Theorem \ref{thm_DOSoM_main} for ESO. We note that the ergodicity of the family $\mathcal{T}$ and the covariance of $H_\omega$ insure the existence of the DOSm: There exists a set $\Omega_0 \subset  \Omega$, with $\Pp ( \Omega_0) = 1$, such that for all $\omega \in \Omega_0$ and for all $x \in \mathcal{V}$, the local DOSoM $n_{V_\omega;x}^*$ equals the 
DOSm which is independent of $x \in \mathcal{V}$. That is, we have that for all $\omega \in \Omega_0$,  for all $f \in \mathcal{C}_c (\R)$, and for all $x \in \mathcal{V}$,  
\beq\label{eq:DefnDOSergodic1}
n_{V_\omega;x}^* (f) = \E \{  \langle \delta_x, f(H_\omega) \delta_x  \rangle  \} =: n_v (f) ,
\eeq
where the potential $V_\omega$ and the sampling function $v$ are related by \eqref{eq:ErgodicPot1}. This also serves to define the DOSm $n_v$ for an ESO with sampling function $v$. 
As pointed out by Bourgain and Klein \cite[equation (1.12)]{bourgain-klein1}, in general, the  DOSm (if it exists) is only bounded above by the DOSoM.  

In the next theorem, we present our main result on the modulus of continuity of the DOSm for ESO on an infinite graph satisfying the UGH.
We recall that $\sigma( \Delta_\mathbb{G})  = [ - \rho_{\mathbb{G}}, \rho_{\mathbb{G}} ]$, for some finite $\rho_{\mathbb{G}}> 0$. 

\begin{theorem} \label{thm_DOSoM_ESOmain}
Let $\mathbb{G}=(\mathcal{V},\mathcal{E})$ be an infinite, connected graph satisfying the UGH and let $H_\omega$ be a family of ergodic Schr\"odinger operators  with probability space $(\Omega, \mathcal{B}, \Pp)$  and with a sampling function $v \in  L^\infty( ( \Omega, \Pp); \R) $, taking values in $[-C, C]$, for some finite $C > 0$. 
Then, the map 
\begin{equation} \label{eq_mainthmmap2}
\mathcal{N} : L^\infty( ( \Omega, \Pp); [-C,C]) \to \mathcal{SL}(\mathcal{C}([-\rho_\mathbb{G} - C, \rho_\mathbb{G} + C])) ~\mbox{, } v \mapsto n_v
\end{equation}
is continuous with respect to the topology for the codomain induced by (\ref{eq_weakconv_metric}).
 The modulus of continuity of the DOSm is quantified by the following: if $\gamma_\mathbb{G}$ is a function satisfying (\ref{eq_growthfunction}), then for every $\zeta >0$, and any pairs of sampling functions $v, u \in L^\infty ((\Omega, \Pp); [-C,C])$, one has 
\begin{equation} \label{eq_mainthm2}
d_w(n_v,  n_u) \leq 2^{3/2} (\rho_\mathbb{G} + C) c_b \cdot \dfrac{1}{\sqrt{  \gamma_\mathbb{G}^{-1}\left(\left(\dfrac{1}{\Vert v - u \Vert_\infty}\right)^\zeta \right) }    } +  \left( \Vert v - u \Vert_\infty \right)^{1 - \zeta} .
\end{equation}
\end{theorem}


\subsubsection{Example 1: $\mathbb{G} = \Z^d$}

 Let $(\Omega, \mathcal{B}, \Pp)$ be a measure space with a family $\mathcal{T} = \{ T_j \}_{j=1}^d$ of invertible ergodic maps $T_j: \Omega \rightarrow \Omega$ providing a representation of the commutative, additive group $\Z^d$. 
For fixed $C > 0$, we construct a potential $V_\omega$ on $\Z^d$ using a \emph{sample function } $v \in L^\infty (( \Omega, \Pp); [-C,C])$ by
\beq\label{eq:ErgPot1}
V_\omega (n) := v( \Pi_{j=1}^d T_j^{n_j} \omega), ~~~n = ( n_1, \ldots, n_d) \in \Z^d ~~\rm{and}~~\omega \in \Omega.
\eeq
Taking $L_{\Z^d} = \Delta_\mathbb{G}$, the standard finite-difference Laplacian on $\Z^d$, the ESO is
$H_\omega = \Delta_{\Z^d} + V_\omega$.

\begin{theorem} \label{thm_ergodic_lattice_DOSoM}
Consider the $d$-dimensional lattice $\mathbb{G} = \mathbb{Z}^d$, $d \in \mathbb{N}$. Then, for each fixed $C>0$, the map in (\ref{eq_mainthmmap2}) is Lipschitz continuous with respect to the topology for the codomain induced by the pseudometric in (\ref{eq_weakconv_metric}), i.e.\  for all sampling functions $v, u \in L^\infty ((\Omega, \Pp); [-C,C])$, with $\Vert u-v \Vert_\infty < 1$, one has
\begin{equation} \label{eq_ergodic_mainthm_zd}
d_w(n_u, n_v) \leq \Vert u - v \Vert_\infty ~\mbox{.}
\end{equation}
\end{theorem}

\subsubsection{Example 2: Bethe lattice $\mathbb{B}_k$}

Let $\mathbb{B}_k$ be a Bethe lattice with coordination number $k$, for $k \geq 3$. This means that each vertex has $k$ nearest neighbors.  A family of ergodic maps $\mathcal{T}$ for the Bethe lattice $\mathbb{B}_k$ was introduced by Acosta and Klein \cite[Appendix]{AcostaKlein}. As in this appendix, we choose one vertex and label it as the origin $(0)$. The origin has $k$ nearest neighbors and this set of vertices constitutes the first level. The vertices of the first level are labeled by one integer $(a_1)$, where $a_1 = 1,2, \ldots, k$. Level 2 consists of those vertices that are nearest-neighbors to the vertices of level 1 and not previously labeled. These are labeled  by a pair of integers $(a_1, a_2)$ designating the vertex connected to $0$ passing through $(a_1)$. Similarly, on level $\ell$, the vertices are labeled $(a_1, a_2, \ldots, a_\ell)$, with $a_1 =1, 2, \ldots, k$ and $a_j = 1, 2, \ldots, k-1$, for $j=2, \ldots, \ell$. This gives a radial structure to $\mathbb{B}_k$ for which the distance from the origin is given by the level number $\ell$. 

Acosta and Klein define two automorphisms of $\mathbb{B}_k$, $\tau_j$, for $j=1,2$. The first automorphism $\tau_1$ is a generalized translation:
\begin{align}\label{eq:transl1}
 &\tau_1 (0)  =   (1) \nonumber \\
& \tau_1 (k) = (0) \nonumber \\
 &\tau_1 (a_1, \ldots, a_\ell)  =   (1, a_1, \ldots, a_\ell) , ~~{\rm if} ~~  \ell \geq 2, ~~\mbox{\rm and} ~~1  \leq  a_1 \leq k-1 \nonumber \\
  &\tau_1 (k, a_2,  \ldots, a_\ell) =   (a_2+1, a_3,  \ldots, a_\ell) , ~~{\rm if} \ell \geq 2, ~~\mbox{\rm and} ~~  a_1 = k .
\end{align}
For $\ell \geq 2$, the effect of $\tau_1$ is to move vertices from level $\ell$ to level $\ell + 1$, followed by a reorientation. The second automorphism $\tau_2$ is a generalized rotation:
 \begin{align}\label{eq:transl2}
 & \tau_2 (0)  =   ( 0)  \nonumber \\
  & \tau_2 (a_1, \ldots, a_\ell)  =    ( ( a_1 + 1){\rm mod} (k), (a_2 + 1) {\rm mod} (k-1) ,  \ldots, (a_\ell + 1) {\rm mod} (k-1)) .
 \end{align}
 The map $\tau_2$ preserves the level index and rotates vertices about each vertex on a given level. 

The main characteristics of the maps $\{ \tau_1, \tau_2 \}$ are
\begin{enumerate}
\item  They act transitively on $\mathbb{B}_k$: For any $x \in \mathbb{B}_k$, there are indices $d_1, d_2 \in \N$ so that 
\beq\label{eq:transitive1}
x = \tau_2^{d_2} \tau_1^{d_1} (0) ;
\eeq

\item The maps do not commute ;

\item They are invertible.

\end{enumerate}
 These maps play the role of the $\{ \alpha_x \}$ described above where, by \eqref{eq:transitive1}, the pairs of indices $(d_1, d_2)$ label the vertices and hence the maps 
$\{ \alpha_{d_2,d_1} ~|~ d_1, d_2 \in \Z \}$.

The probability space $(\Omega , \mathcal{B}, \Pp)$ is  assumed to have the property that the maps $\{ \alpha_{d_2,d_1} ~|~ d_1, d_2 \in \Z \}$ induce an action on $\Omega$ forming the family of maps $\mathcal{T} = \{ T_{d_2, d_1} ~|~ d_1, d_2 \in \Z \}$.
We assume that this family $\mathcal{T}$ is an ergodic family of invertible maps on the probability space $(\Omega , \mathcal{B}, \Pp)$. Acosta and Klein \cite{AcostaKlein} considered the case when $(\Omega, \mathcal{B}, \Pp)$ is a product probability space. In this case, we have
we have  
\beq\label{eq:probMaps1}
T_{d_2, d_1} \omega = ( \omega_{\alpha_{d_2, d_1}( x)})_{x \in \mathbb{B}_k} .
\eeq
In this setting, the map $\tau_1^{d_1}$ is ergodic so  the family $\mathcal{T}$ is ergodic.


We construct an ESO as follows. Given a sampling function $v \in L^\infty (( \Omega, \Pp);[C, -C])$,for some $C > 0$, we define a potential $V_\omega (x)$, for $x \in \mathbb{B}_k$, with $x = \tau_2^{d_2} \tau_1^{d_1}(0)$,  by
\beq\label{eq:potentialBethe1}
V_\omega (x) := v(T_{{d_2}, {d_1}} \omega)  , ~~ \omega \in \Omega .
\eeq
The discrete Laplacian $\Delta_{\mathbb{B}_k}$ is invariant under the transformations on $\ell^2 (\mathbb{B}_k)$ induced by $\tau_j$, for $j=1,2$. 
Consequently, the Schr\"odinger operator  $H_\omega := \Delta_{\mathbb{B}_k} + V_\omega$ is an ESO. We recall that the spectral radius of the Laplacian $\Delta_{\mathbb{B}_k}$ is $\rho_{\mathbb{B}_k} = 2 \sqrt{k-1}$.

\begin{theorem} \label{thm_ergodic_Bethe_DOS}
Consider the Bethe lattice $\mathbb{G} = \mathbb{B}_k$  with coordination number $k \geq 3$, and affiliated probability space $( \Omega, \mathcal{B}, \Pp)$ as described above. 
 Then, for each $C > 0$, the map in \eqref{eq_mainthmmap} is $\frac{1}{2}-\log$-H\"older continuous with respect to the topology for the codomain induced by the pseudometric in (\ref{eq_weakconv_metric}), i.e. there exists a constant $\gamma_k > 0$, explicitly given in (\ref{eq_bethe_constant}), such that for all sampling functions $u, v \in L^\infty((\Omega, \mu); [-C,C])$ with $\Vert v - u  \Vert_\infty < 1$, one has
\begin{equation} \label{eq_thm_ergodic_Bethe_modulus}
d_w(n_v^*, n_u^*) \leq \dfrac{\gamma_k}{\sqrt{\log\left(\frac{1}{ \Vert v - u \Vert_\infty}    \right)     }} ~\mbox{.}
\end{equation}
\end{theorem}

%
%
%
%
%

\subsection{Random Schr\"odinger operators}\label{subsec:RSO1}

We recall the usual formulation of a random Schr\"odinger operator (RSO) on a graph $\mathbb{G}$. Let $L_{\mathbb{G}}$ be a discrete finite difference operator on $\mathbb{G}$, like the Laplacian or the adjacency matrix. Let $\mu$ denote a single-site probability measure with support in $[-C, C]$, for some $0 < C < \infty$. We denote by $\Omega$ the product probability space $\Omega := [-C, C]^\mathcal{V}$ and the product probability measure $\mathbb{P}_\mu = \otimes_{x \in \mathcal{V}} ~ \mu $.
We form the probability space $(\Omega, \mathcal{B}, \mathbb{P}_\mu)$ and write $\omega = ( \omega_x)_{x \in \mathcal{V}} \in \Omega$. The Anderson potential on $\mathbb{G}$ is given by
\beq\label{eq:AndPot1}
( V_\omega \psi)(x) = \omega_x \psi (x) , ~~ \psi \in \ell^2 ( \mathbb{G}; \mathbb{C}) ,
\eeq
and the corresponding RSO is
\beq\label{eq:RSO1}
H_\omega := \Delta_{\mathbb{G}} + V_\omega .
\eeq

For a single-site probability measure $\mu$ with support in $[-C,C]$, the cumulative distribution function (CDF) is denoted by $F_\mu: \R \rightarrow [0, 1]$. We define the corresponding \emph{quantile function }$q_\mu : [0,1] \rightarrow [-C,C]$ by
\beq\label{eq:quantile1}
q_\mu (x) := \inf_{t \in \R} \{ F_\mu (t) \geq x \} .
\eeq
We note that if the CDF $F_\mu$ is strictly increasing on ${\rm supp} ~\mu$, then the corresponding quantile function is given by $q_\mu = F_\mu^{-1}$. 

We recal some properties about the convergence of probability measures, their CDF, and the corresponding quantile functions. We refer the reader to \cite{parzen} for more information on quantile functions and their relation to the CDF.

For every probability measure $\mu$ with support in $[-C,C]$, one has
\begin{equation} \label{eq_representation}
\mu = \mu_L \circ q_\mu^{-1} ~\mbox{,}
\end{equation}
where $\mu_L$ is Lebesgue measure on $[0,1]$ and $q_\mu^{-1}$ denotes the pre-image map associated with $q_\mu$. 

Let $\mu$ and a sequence $\{ \mu_n \}$ be probability measures all having support on a fixed set $[-C,C]$. It is well-known (see e.g. \cite[page 153]{rachev}) that the weak convergence of measures $\mu_n \rightarrow \mu  ~~{weakly}$ is equivalent to the convergence of the quantile functions $q_{\mu_n} \rightarrow q_\mu$ in $L^1 ( [0,1]; \mu_L)$. Indeed, $L^1$-convergence of the quantile functions may be expressed in terms of the Kantorovich-Rubinstein-Wasserstein (KRW) metric 
\beq\label{eq:KRWmetric1}
d_{\rm KRW} (\mu, \nu)  := \|    q_\mu- q_\nu \|_{L^1 ([0,1]; \mu_L)} ~\mbox{.}
\eeq
The more common expression of the KRW metric from optimal transport theory, which was also used in \cite{shamis}, coincides with the expression as $L^1$-distance of the quantile functions given here; see e.g. \cite[page 153]{rachev}.

The equivalence between the KRW metric (\ref{eq:KRWmetric1}) and the metric in \eqref{eq_weakconv_metric} is expressed by the following relation: given two probability measure $\mu, \nu$ with support in $[-C, C]$, one has 
\begin{equation} \label{eq_equalitymetrics}
d_w (\mu, \nu) \leq d_{\mathrm{KRW}} (\mu, \nu ) \leq (1 + C) d_w(\mu, \nu) ~\mbox{.}
\end{equation}
Since we could not find an explicit proof of (\ref{eq_equalitymetrics}) in the literature, we include a brief argument in Appendix \ref{app:equivmetric}.

%
%
%

In terms of the quantile function $q_\mu$, the Anderson potential \eqref{eq:AndPot1} may be expressed as
\beq\label{eq:quantile2}
V_\omega (x) = q_\mu (\tilde{\omega}_x) , ~~ {\rm for} ~~ \tilde{\omega}_x \in [0, 1]^\mathcal{V}, 
\eeq
and the random variables $\tilde{\omega}_x$, $x \in \mathcal{V}$, are uniformly distributed on $[0,1]$.
From the definition of the quantile function $q_\mu$ in \eqref{eq:quantile1} and (\ref{eq_representation}), it follows that the two random variables
\beq\label{eq:RV1}
V_\omega (x) = \omega_x, ~~{\rm on } ~~ ([-C,C], \mathcal{B}_C, \mu) , 
\eeq
and
\beq\label{eq:RV2}
q_\mu (\tilde{\omega}_x) ,  ~~{\rm on } ~~ ([0,1],  \mathcal{B}_1, \mu_L) , 
\eeq
are equal in distribution. Here, $\mathcal{B}_C$, respectively, $\mathcal{B}_1$, is the set of  Borel subsets of $[-C,C]$, respectively, of $[0,1]$.

The advantage of this reformulation of the Anderson model in terms of the quantile function is that this representation explicitly shows that a  RSO is an ESO as defined in section \ref{subsec:ergodicSO1} where the dependence on the single-site probability measure only enters through the potential \eqref{eq:quantile2}. In particular, the dependence of the DOSm for a  RSO on the single-site probability measure $\mu$ becomes a special case of the dependence of the DOSm for ESO on the sampling function in
Theorem \ref{thm_DOSoM_ESOmain}, in particular,    
 Theorem \ref{thm_ergodic_lattice_DOSoM} for the lattice, and Theorem \ref{thm_ergodic_Bethe_DOS} for the Bethe lattice. To see this, we construct a probability space $([0,1]^\mathcal{V}, \mathcal{B}, \Pp_L)$, where $\Pp_L$ is the product measure
$\Pp_L := \otimes_{x \in \mathcal{V}} \mu_L$, and where $\mu_L$ is the Lebesgue measure on $[0,1]$. Taking $q_\mu$ as the sampling function, the operator $H_{\tilde{\omega}} = \Delta_\mathbb{{G}} +  V_{\tilde{\omega}}$ is an ESO on $\mathbb{G}$.


Using the quantile functions $q_\mu$ and $q_\nu$ associated with two probability measure $\mu$ and $\nu$ on $[-C, C]$, the inequality in Lemma \ref{lemma_keyfinitereduction} becomes the following statement:
\begin{align}
\left\vert  \dfrac{1}{\vert \Lambda_L^{(\mathbb{G})}(x) \vert} \right.  &   \left.  \left[ \mathrm{Tr} \left( P_L^{(\mathbb{G})}(x) f(H_W) P_L^{(\mathbb{G})}(x) \right) - \mathrm{Tr} \left( P_L^{(\mathbb{G})}(x) f(H_{V_W^{(M;x)}})  P_L^{(\mathbb{G})}(x) \right) \right]   \right\vert   \nonumber  \\ 
& \leq  \dfrac{ L_f}{\vert \Lambda_L^{(\mathbb{G})}(x) \vert}   \sum_{y \in \Lambda_M^{(\mathbb{G})} (x)}
| \omega_y(\mu) - \omega_y(\nu) | 
~\mbox{.} \label{eq_lemma_lipschitz_3}
\end{align}
We now take the expectation $\E_L$ with respect to the product measure $\mathbb{P}_L := \otimes_{x \in \mathcal{V}}  \mu_L$:
\beq\label{Eq:RSOexp1}
\E_L \left\{  \sum_{y \in \Lambda_M^{(\mathbb{G})} (x)}| \omega_y(\mu) - \omega_y(\nu) |  \right\}
= \| q_\mu - q_\nu \|_{L^1 ([0,1], \mu_L)} ~| \Lambda_M^{(\mathbb{G})}(x)| .
\eeq
 Consequently, we obtain an upper bound of the left side of \eqref{eq_lemma_lipschitz_3}:
\beq\label{eq:finitereduction1q}
  L_f  \dfrac{| \Lambda_M^{(\mathbb{G})}(x) |}{\vert \Lambda_L^{(\mathbb{G})}(x) \vert}  ~\| q_\mu - q_\nu \|_{L^1([0,1], \mu_L)} .
\eeq

With these modifications, the results for ESO in Thoerem \ref{thm_DOSoM_ESOmain}, Theorem \ref{thm_ergodic_lattice_DOSoM} for the lattice, and Theorem \ref{thm_ergodic_Bethe_DOS} for the Bethe lattice, provide the following quantitative bounds on the modulus of continuity for the DOSm improving the results of \cite[Theorem 3.1, Theorem 6.1]{hislop_marx_1}.
In the following, we let $\mathcal{P}([-C, C])$ denote the space of Borel probability measures on $[-C,C]$. 
 
\begin{theorem} \label{thm_ergodic_random_DOS}
We consider a RSO $H_\omega$ as in \eqref{eq:AndPot1} and \eqref{eq:RSO1} on $\ell^2 ( \mathbb{G} ; \mathbb{C})$. 
Then, for each fixed $C>0$, the map 
\begin{equation} \label{eq_mainthmmap22}
\mathcal{N} : (\mathcal{P} ( [-C, C]), d_{\mathrm{KRW}} )  \to (\mathcal{P}([-\rho_\mathbb{G} - C, \rho_\mathbb{G} + C]), d_w) ~\mbox{, } \mu \mapsto n_\mu
\end{equation}
is continuous.  
 The modulus of continuity of the DOSm is quantified by the following: If $\gamma_\mathbb{G}$ is a function satisfying (\ref{eq_growthfunction}), then for every $\zeta >0$, and any pair of  probability measures 
$\mu, \nu \in \mathcal{P}([-C,C])$,
\begin{equation} \label{eq_mainthm3}
d_w(n_\nu,  n_\mu) \leq 2^{3/2} (\rho_\mathbb{G} + C) c_b \cdot \dfrac{1}{\sqrt{  \gamma_\mathbb{G}^{-1}\left(\left(\dfrac{1}{ d_{\mathrm{KRW}}(\mu , \nu) }\right)^\zeta \right) }    } +  \left( d_{\mathrm{KRW}} (\mu, \nu) \right)^{1 - \zeta} .
\end{equation}


\noindent
In particular, we have 
\begin{itemize}
\item[(i)] For the $d$-dimensional lattice $\mathbb{G} = \mathbb{Z}^d$, $d \in \mathbb{N}$, or any lattice satisfying the UGH with a uniform growth function comparable to one for $\Z^d$,
we have
\begin{equation} \label{eq_ergodic_random_mainthm_zd}
d_w (n_\mu, n_\nu) \leq d_{\rm KRW} (\mu, \nu)  ~\mbox{.}
\end{equation}

\item[(ii)] For the Bethe lattice $\mathbb{G} = \mathbb{B}_k$  with coordination number $k$, we have
\begin{equation} \label{eq_thm_ergodic_random_Bethe_modulus}
d_w (n_\mu, n_\nu) \leq \dfrac{\gamma_k}{\sqrt{\log\left(\frac{1}{ d_{\rm KRW}(\mu, \nu) }    \right)     }} ~\mbox{,}
\end{equation}
where the constant $\gamma_k$ is defined in  \eqref{eq_bethe_constant}.

\end{itemize}
\end{theorem}



\begin{appendices}

\section{Appendix: Definitions of the DOSoM for lattice Schr\"odinger operators} \label{app_DOSoM_lattice}
\setcounter{equation}{0}

In this section, we show that the definitions of the DOSoM (\ref{eq_defnDOSoM}), as well as the modulus of continuity for its cumulative distribution, the IoDS in (\ref{def_IDS}), given in this note for general graphs $\mathbb{G}$, coincide with the respective quantities used by Bourgain and Klein \cite{bourgain-klein1} for the special case that $\mathbb{G} = \mathbb{Z}^d$. We recall that in definition \eqref{eq_defnDOSoM} we use the infinite volume operator $H_V$ whereas  Bourgain- Klein \cite{bourgain-klein1}  use the restriction of the operator to finite subsets.  

As shown in  (\ref{eq_compareBKdefn}), the key is essentially the second resolvent identity. For the case of $\mathbb{G} = \mathbb{Z}^d$, this identity implies that the error term between the two definitions of the DOSoM decays to zero. However, 
since the DOSoM is an {\em{outer}} measure, it is, in general, only subadditive, as opposed to additive. The purpose of this section is to provide the details, accounting for this minor technicality, relating the two definitions for the case $\mathbb{G} = \mathbb{Z}^d$.

To do so, we recall that $H_{V;L}(x)$ is the finite-volume restriction of the operator $H_V$ to $\Lambda_L(x)$ defined in (\ref{eq_finitevolumerestr}). We first introduce the local DOSoM according to Bourgain-Klein \cite{bourgain-klein1} given by,
\begin{align} \label{eq_defnDOSoM_local_BK}
n_{V;x}^{*;\mathrm{BK}}(f):= \limsup_{L \to \infty} \left\{ \dfrac{1}{\vert \Lambda_L^{(\mathbb{Z}^d)}(x) \vert} \mathrm{Tr}\left( P_L^{(\mathbb{Z}^d)}(x) f(H_{V;L}(x)) P_L^{(\mathbb{Z}^d)}(x) \right)  \right\} ~\mbox{,} 
\end{align}
and its associated cumulative distribution, the local IoDS, 
\begin{equation} \label{def_IDS_BK}
N_{V;x}^{*;\mathrm{BK} }(E):= n_{V;x}^{*;\mathrm{BK}}(\chi_{(-\infty ,E]}) ~\mbox{, for $E \in \mathbb{R}$ .  }
\end{equation}
 It was shown in \cite{bourgain-klein1} that for each $x \in \mathbb{Z}^d$ and $a < b$ with $0 < b - a < \frac{1}{2}$ one has 
\begin{equation} \label{eq_BK_app}
n_{V;x}^{*;\mathrm{BK}}(\chi_{[a ,b]}) \leq \dfrac{K_{d;V}}{\log(\frac{1}{b-a} )} ~\mbox{,}
\end{equation}
for some constant $K_{d; V}$ only depending on $d$ and $\Vert V \Vert_\infty$ and independent of $x \in \mathcal{V}$.

We observe that by the continuous functional calculus, the density in $\mathcal{C}_0(\mathbb{R})$ of the algebra generated by $\{f_a ~|~ a \in \mathbb{C} \setminus \mathbb{R}\}$, where $f_a$ is defined in (\ref{eq_2ndresolvent_decay}), implies that (\ref{eq_compareBKdefn}) extends to all $f \in \mathcal{C}_0(\mathbb{R})$. Here, as common, $\mathcal{C}_0(\mathbb{R})$ denotes the continuous functions vanishing at infinity equipped with the supremum norm. Thus, for every $f \in \mathcal{C}_0(\mathbb{R})$, the two definitions of the DOSoM agree, i.e.
\begin{equation} \label{eq_app_DOSoM_agree_1}
n_{V;x}^{*;\mathrm{BK}}(f) = n_{V;x}^{*}(f) ~\mbox{.}
\end{equation}

We conclude by demonstrating that local IoDS defined in this note
\begin{equation} \label{eq:dosom2}
N_{V;x}^*(E):= n_{V;x}^{*}(\chi_{[a ,b]}) ~\mbox{,}
\end{equation}
admits the same modulus of continuity for $\mathbb{G} = \mathbb{Z}^d$ as stated in (\ref{eq_BK_app}). Morally, we thus aim to extend (\ref{eq_app_DOSoM_agree_1}) to the functions $f = \chi_{[a,b]}$, for $a < b$ with $b - a < \frac{1}{2}$. Since the objects involved in (\ref{eq_app_DOSoM_agree_1}) are only outer measures, we take advantage of the monotonicity of  \eqref{eq_defnDOSoM_local_BK} and \eqref{eq:localDOSoM4} in the function $f$. For $\epsilon > 0$, let $0 \leq \phi_{\epsilon} \leq 1$ be continuous with $\mathrm{supp} ~\phi_{\epsilon} = [a-\epsilon, b + \epsilon]$ and $\phi_\epsilon =1$ on $[a,b]$. Then, the monotonicity described above, (\ref{eq_app_DOSoM_agree_1}), and (\ref{eq_BK_app}) imply that
\begin{align}
n_{V;x}^{*}([a,b]) & \leq n_{V;x}^{*}(\phi_\epsilon) = n_{V;x}^{*;\mathrm{BK}}(\phi_\epsilon) \leq n_{V;x}^{*;\mathrm{BK}}([a - \epsilon , b + \epsilon]) \nonumber \\ 
  & \leq \dfrac{K_{d;V}}{\log \left( \frac{1}{b-a + 2 \epsilon} \right) } ~\mbox{,} \label{eq_app_showequ}
\end{align}
whence, taking $\epsilon \to 0^+$ on the right-most side of (\ref{eq_app_showequ}) yields
\begin{equation}
n_{V;x}^{*}([a,b]) \leq \dfrac{K_{d;V}}{\log \left( \frac{1}{b-a} \right) } ~\mbox{,}
\end{equation}
which agrees with (\ref{eq_BK_app}), as claimed.



\section{Appendix: An alternate proof based on results of Aleksandrov and Peller }\label{app:AltProofAP1}

\setcounter{equation}{0}

I.\  Kachkovskiy \cite{kachkovskiy} indicated to us the following result of Aleksandrov and Peller \cite{aleksandrovPeller2011} which provides an alternate proof of some of the results in our paper.  For a Lipschitz function $f$ on $\R$, the Lipschitz constant $L_f$ is defined in \eqref{eq_lipschitznorm}.

\begin{theorem} \label{thm:peller1}
Let $f \in {{\rm Lip}} (\R)$. Let $A$ be a bounded self-adjoint operator with $\sigma(A) \subset [a, b]$. Then for
every bounded self-adjoint operator $B$,
\beq \label{diffAP1}
\| f(A) - f(B) \| \leq C_0 ~ L_f ~\log \left(   2  + \frac{b-a}{\| A - B \|} \right) ~ \| A - B \| ,
\eeq
for some universal finite constant $C_0>0$.
\end{theorem}

Although Theorem \ref{thm:peller1} is a general result, the proof is rather involved and does not give the optimal modulus of continuity of the DOSoM for lattices satisfying the UGH for which the growth function is bounded relative to that for $\mathbb{G} = \Z^d$. The proof presented in our paper is self-contained and rather elementary.  Theorem \ref{thm:peller1} does, however, allow for an improvement of our results for the case of the Bethe lattice.

Theorem \ref{thm:peller1}  implies the  following analog of Lemma \ref{lemma_keyfinitereduction}.
on the modulus of continuity of the DOSoM for Schr\"odinger operators on graphs $\mathbb{G}$ with the UGH. Rather than applying Proposition \ref{prop_lipschitz_traceclass} to prove Lemma  \ref{lemma_keyfinitereduction}, we apply Theorem \ref{thm:peller1} in order to estimate the difference of the traces on the left side of \eqref{eq_lemma_lipschitz_AP4} directly. We note that the trace is controlled by the trace class operator $P_L^{(\mathbb{G})} (x)$. 

\begin{corollary}\label{lemma_ap1}
Let $x \in \mathcal{V}$ and $M > 0$ be fixed. Then, for all $V,W \in \ell^\infty(\mathcal{V}; [-C, C])$ and $L > 0$, one has
\begin{align}\label{eq_lemma_lipschitz_AP4}
\left\vert \dfrac{1}{\vert \Lambda_L^{(\mathbb{G})}(x) \vert} \right. & \left. \left[ \mathrm{Tr} \left( P_L^{(\mathbb{G})}(x) f(H_W) P_L^{(\mathbb{G})}(x) \right) - \mathrm{Tr} \left( P_L^{(\mathbb{G})}(x) f(H_{V})  P_L^{(\mathbb{G})}(x) \right) \right] \right\vert  \nonumber  \\ 
& \leq C_0 ~  L_f ~ \log \left( 2 + \frac{2 (\rho_{\mathbb{G}}+ C)}{ \Vert V - W \Vert_\infty } \right) ~   
\Vert V - W \Vert_\infty ,  
\end{align}
where the constant $C_0$ is the same as in \eqref{diffAP1}. 
\end{corollary}


With this estimate, we obtain  the following version of Theorem \ref{thm_DOSoM_main}:

\begin{theorem} \label{thm_DOSoM_mainV2}
Let $\mathbb{G}=(\mathcal{V},\mathcal{E})$ be an infinite, connected graph. 
Then, for each finite $C>0$, the map
\begin{equation} \label{eq_mainthmmapAP}
\mathcal{N}_C: \ell^\infty(\mathcal{V};[-C,C]) \to \mathcal{SL}(\mathcal{C}([-\rho_\mathbb{G} - C, \rho_\mathbb{G} + C])) ~\mbox{, } V \mapsto n_V^*
\end{equation}
is continuous with respect to the topology for the codomain induced by (\ref{eq_weakconv_metric}). The modulus of continuity is quantified by the following: 
\begin{equation} \label{eq_mainthmV2}
d_w(n_V^*, n_W^*) \leq  C_0 ~ \log \left( 2 + \frac{2 ( \rho_{\mathbb{G}}+ C)}{ \Vert V - W \Vert_\infty } \right) ~   \Vert V - W \Vert_\infty .
\end{equation}
for all $V, W \in \ell^\infty(\mathcal{V};[-C,C])$ with $\Vert V - W \Vert_\infty < 1$. Here,  the constant $C_0$ is the same as in \eqref{diffAP1}.
\end{theorem}


To compare this result with Theorem \ref{thm_DOSoM_main}, we look at the application to  $\mathbb{G} = \Z^d$ and to $\mathbb{G} = \mathbb{B}_k$ in section 
\ref{sec:applications1}.  For the square lattice $\mathbb{G} = \Z^d$, and under the hypothesis of Theorem \ref{thm_DOSoM_main}, we obtain the Lipschitz continuity of the DOSoM:
\beq\label{eq:DOSoMcont1}
d_w ( n_V^* , n_W^* ) \leq \| V - W \|_\infty,
\eeq
which is a stronger continuity result than the application of Theorem \ref{thm_DOSoM_mainV2} to the lattice in \eqref{eq_mainthmV2} with $\rho_{\mathbb{G}} = 2d$. 
With regards to the Bethe lattice $\mathbb{G} = \mathbb{B}_k$, result \eqref{eq_mainthmV2}, with $\rho_{\mathbb{B}_k} =  2\sqrt{k-1}$ is a stronger results that
the one obtained by the methods of this paper
\begin{equation} \label{eq_thm_Bethe_modulus2}
d_w(n_V^*, n_W^*) \leq \dfrac{\gamma_k}{\sqrt{\log\left(\frac{1}{ \Vert V - W \Vert_\infty}    \right)     }} ~\mbox{,}
\end{equation}
where $\gamma_k$ is defined in \eqref{eq_bethe_constant}.

\section{Appendix: Proof of the inequality (\ref{eq_equalitymetrics}) between the KRW-metric $d_{\mathrm{KRW}}$ and the Fortet-Mourier metric $d_w$}\label{app:equivmetric}
\setcounter{equation}{0}

In this appendix, we supply a proof of the equivalence of the metrics $d_{\mathrm{KRW}}$ and $d_w$ as expressed by the inequality in (\ref{eq_equalitymetrics}). As in section \ref{subsec:RSO1}, for given $0 < C < +\infty$, we let $\mathcal{P} ( [-C, C])$ denote the space of Borel probability measures on $[-C,C]$.

The leftmost inequality in (\ref{eq_equalitymetrics}),
\begin{equation} \label{eq_equalitymetrics_leftmost}
d_w (\mu, \nu) \leq d_{\mathrm{KRW}}(\mu, \nu) ~\mbox{, } \mu, \nu \in \mathcal{P} ( [-C, C]) ~\mbox{,}
\end{equation}
is an immediate consequence of (\ref{eq_representation}), since for every $f \in \mathrm{Lip}([-C,C])$, one has
\begin{align}
\left\vert \int_{-C}^C f(x) ~\mathrm{d} \mu(x) - \int_{-C}^C  f(x) ~\mathrm{d} \nu(x) \right\vert & = \left\vert \int_0^1  \left( f(q_\mu(t)) - f(q_\nu(t)) \right) ~\mathrm{d} t \right\vert \nonumber \\
  & \leq L_f \Vert q_\mu - q_\nu \Vert_{L^1([0,1]; \mu_L)} ~\mbox{.}
\end{align}
Here, as in section \ref{subsec:RSO1}, we use $\mu_L$ to denote the Lebesgue measure on $[0,1]$.

To prove the other inequality in  (\ref{eq_equalitymetrics}), that is, 
\begin{equation} \label{eq_equivmetrics_proof}
d_{\mathrm{KRW}} (\mu, \nu ) \leq (1 + C) ~d_w(\mu, \nu) ~\mbox{,}
\end{equation} 
we note that (\ref{eq_equivmetrics_proof}) follows trivially if $q_\nu \leq q_\mu$ on $[0,1]$. Indeed, in this case, the representation (\ref{eq_representation}) of $d_{KRW}$ and the definition of $d_w$ in \eqref{eq_weakconv_metric} imply
\begin{align} \label{eq_trivialcase}
d_{\mathrm{KRW}} (\mu, \nu ) & = \int_0^1 \left( q_\nu(t) - q_\mu(t) \right) \mathrm{d} t  \nonumber = \int_{-C}^C x ~\mathrm{d} \nu(x) - \int_{-C}^C x ~\mathrm{d} \mu(x) \\
       &  \leq \Vert \mathrm{id}_{[-C,C]} \Vert_{\mathrm{Lip}} \cdot d_w(\mu, \nu) = (1 + C) ~d_w(\mu, \nu) ~\mbox{.}
\end{align}
Here, $\mathrm{id}_{[-C,C]}$ is the identity map on $[-C,C]$.

We now reduce the case for an arbitrary pair of measures $\nu, \mu \in  \mathcal{P}([-C,C])$ to this trivial situation. We first observe that it suffices to consider measures in the class
\begin{equation}
\mathcal{P}_0([-C,C]):= \left\{ \mu \in \mathcal{P}([-C,C]) ~:~ F_\mu ~\mbox{is continuous and {\em{strictly}} increasing} \right\} ~\mbox{,}
\end{equation}
which forms a dense subset of $\mathcal{P}([-C,C])$ in the weak topology. After removing atoms using approximate delta functions, every continuous increasing function $F: [-C,C] \to [0,1]$ can be uniformly approximated by a sequence of continuous and {\em{strictly}} increasing functions with the same domain and codomain as $F$, which, by Helly's Second Selection Theorem, implies weak convergence of the associated measures. The key property of $\mathcal{P}_0([-C,C])$ is that $\mu \in \mathcal{P}_0([-C,C])$ if and only if {\em{both}} $F_\mu$ and $q_\mu$ are continuous and strictly increasing. 

Given two measures $\mu, \nu \in \mathcal{P}_0([-C,C])$, continuity implies that the set $\{t \in [0,1] : q_\mu - q_\nu \neq 0\}$ is relatively open in $[0,1]$. Therefore, the interval $[0,1]$ is partitioned into a countable set of closed intervals $\{I_k : k \in \mathbb{N}\}$, which are pairwise disjoint except possibly at endpoints, such that 
$$[-C,C] = \cup_{k \in \mathbb{N}} (q_\mu - q_\nu)(I_k) =:\cup_{k \in \mathbb{N}} J_k$$ 
and, for each $k \in \mathbb{N}$,
\begin{equation}
q_\mu \leq q_\nu ~\mbox{ on $I_k$  if and only if } F_\mu \leq F_\nu ~\mbox{ on $J_k$ . }
\end{equation}
In particular, considering the continuous and strictly increasing functions 
\begin{equation}
q_{\mu} \wedge q_{\nu}:= \min\{ q_\mu, q_\nu\} ~\mbox{, } q_{\mu} \vee q_{\nu}:= \max\{ q_\mu, q_\nu\} ~\mbox{,}
\end{equation}
their inverses satisfy
\begin{equation}
(q_{\mu} \wedge q_{\nu})^{-1} = \min\{ F_\mu, F_\nu\} ~\mbox{, } (q_{\mu} \vee q_{\nu})^{-1} = \max\{ F_\mu, F_\nu\}  ~\mbox{.}
\end{equation}
We therefore obtain measures $\mu \wedge \nu ~,~ \mu \vee \nu \in \mathcal{P}_0([-C,C])$ with
\begin{equation}
F_{\mu \wedge \nu } = \min\{ F_\mu, F_\nu\} ~\mbox{and } F_{\mu \vee \nu} = \max\{ F_\mu, F_\nu\}  ~\mbox{,}
\end{equation}
so that 
\begin{align}
q_{\mu \vee \nu}(t) - q_{\mu \wedge \nu}(t) &= \vert q_\mu(t) - q_\nu(t) \vert ~\mbox{, for all $t \in [0,1]$ ,} \\
F_{\mu \vee \nu}(x) - F_{\mu \wedge \nu}(x) &= \vert F_\mu(x) - F_\nu(x) \vert  ~\mbox{, for all $x \in [-C,C]$ ,}
\end{align}
and, for each $k \in \mathbb{N}$ and all intervals $U \subseteq J_k$ , one has
\begin{align} \label{eq_twomeasures}
(\mu \vee \nu)(U) - (\mu \wedge \nu)(U) = \vert \mu(U) - \nu(U) \vert ~\mbox{.} 
\end{align}
Consequently, (\ref{eq_twomeasures}) yields that for each $f \in \mathcal{C}([-C,C])$ , one has
\begin{equation} \label{eq_reduction}
\left\vert \int_{-C}^C  f(x) ~\mathrm{d}\mu(x) - \int_{-C}^C  f(x) ~\mathrm{d} \nu(x) \right\vert = \int_{-C}^C  f(x) ~\mathrm{d} (\mu \vee \nu)(x) - \int_{_C}^C  f(x) ~\mathrm{d} (\mu \wedge \nu)(x) ~\mbox{.}
\end{equation}

In summary, we conclude from (\ref{eq_trivialcase}) and (\ref{eq_reduction}) that
\begin{equation}
d_{\mathrm{KRW}} (\mu, \nu ) = d_{\mathrm{KRW}} (\mu \wedge \nu, \mu \vee \nu) \leq (1 + C) ~d_w(\mu \wedge \nu, \mu \vee \nu) = (1 + C) ~d_w(\mu, \nu) ~\mbox{,}
\end{equation}
which, together with (\ref{eq_equalitymetrics_leftmost}), establishes the claim in (\ref{eq_equalitymetrics}).

\section{Appendix: Continuity of the spectrum in Hausdorff metric} \label{app:spectrumhausdorff}
\setcounter{equation}{0}

In this appendix, we discuss the related question of the continuity of the spectrum of deterministic  Schr\"odinger operators with respect to the potential in the spirit of Theorem \ref{thm_DOSoM_main}. We then present applications  to ESO and RSO and prove results on the continuity of the deterministic spectrum with respect to the sampling function or the single-site probability measure, respectively. 

Theorem \ref{thm_DOSoM_main} presents estimates on the modulus of continuity of the DOSoM with respect to the potential sequence. These estimates yield quantitative continuity results for the DOSm of ESO in the sampling function (Theorem \ref{thm_DOSoM_ESOmain}) and, for RSO, weak continuity of the DOSm in the single-site probability measure (Theorem \ref{thm_ergodic_random_DOS}). For ESO, i.e. $H_\omega$ defined in (\ref{eq:ErgodicPot1})-(\ref{eq_ESO_1}) with underlying  sampling function $v$, the DOSm $n_v$ is related to the almost sure (a.s.)-spectrum $\Sigma_v:=\sigma(H_\omega)$, $\mathbb{P}$-a.s., through the formula 
\begin{equation} \label{eq_asspectrumESO_DOSm}
\mathrm{supp}(n_v) = \Sigma_v ~\mbox{.}
\end{equation}
Thus, it is only natural to ask about the relation between the continuity of the DOSoM in the potential, proven here, and known continuity results of the spectrum in the Hausdorff metric. We recall that for two compact sets $K,L \subseteq \mathbb{R}$, the distance between $K$ and $L$ in the Hausdorff metric is defined by
\begin{equation}
\mathrm{dist}_{\rm H}(K,L):= \max\{ \sup_{x \in K} \mathrm{dist}(x; L) ~;~ \sup_{x \in L} \mathrm{dist}(x; K) \} ~\mbox{.}
\end{equation}
We also note that a useful equivalent formulation of the Hausdorff metric is given in (\ref{eq_hd_equivdef}).

A priori, weak convergence of compactly supported measures does not, of course, imply nor necessitate convergence of their supports in the Hausdorff metric; see also Example \ref{example} below. Basic considerations using the resolvent, see \cite[Theorem 4.10, chapter V, section 4.3]{Kato_book} for details, however show that if $A,B$ are bounded self-adjoint operators then their spectra satisfy
\begin{equation} \label{eq_contispectraHD_basic}
\mathrm{dist}_{\rm H}(\sigma(A),\sigma(B)) \leq \Vert B - A \Vert ~\mbox{.}
\end{equation}
The purpose of this section is to briefly discuss the implications of (\ref{eq_contispectraHD_basic}) for the DOSoM and formulate the associated results for ergodic and random  Schr\"odinger operators using the framework developed in the sections \ref{subsec:ergodicSO1}-\ref{subsec:RSO1}.

We first turn to the deterministic set-up described in section \ref{sec:DOSoM1}. Given a Schr\"odinger operator $H_V$ on a graph $\mathbb{G}=(\mathcal{V}, \mathcal{E})$ as described in (\ref{eq_Schrodop}), for $x \in \mathcal{V}$, consider its 
{\em{local}}  DOSoM $n_{V;x}^*$ at $x$, defined in (\ref{eq:localDOSoM4}). In analogy to measures, we define the support of the {\em{outer}} measure $n_{V;x}^*$ (denoted by $\mathrm{supp}(n_{V;x}^*)$) as the set of $E \in \mathbb{R}$ satisfying
\begin{equation}
\forall \epsilon > 0 ~\mbox{, } n_{V;x}^*(\chi_{(E- \epsilon, E + \epsilon)} ) > 0 ~\mbox{.}
\end{equation}
Note that by monotonicity of outer measures, the support of an outer measure is closed.

In conjunction with  (\ref{eq:localDOSoM4}), this definition of the support of an outer measure implies that 
 if $E \not \in \sigma(H_V)$, then, for some $\epsilon > 0$, $n_{V;x}^*(\chi_{(E- \epsilon, E + \epsilon)} ) = 0$, whence
\begin{equation} \label{eq_spectrum_local_DOSoM}
\mathrm{supp}(n_{V;x}^*) \subseteq \sigma(H_V) ~\mbox{.}
\end{equation}
This is a weakened version of (\ref{eq_asspectrumESO_DOSm}), but it is valid for all deterministic Schr\"odinger operators. The definition of the DOSoM in (\ref{eq_defnDOSoM}) similarly shows that 
\begin{equation}  \label{eq_spectrum_DOSoM}
\mathrm{supp}(n_{V}^*) \subseteq \sigma(H_V) ~\mbox{.}
\end{equation}

We recall that for ESO, the $\mathbb{P}-a.e$ reverse set-inclusion for (\ref{eq_spectrum_local_DOSoM}) is a consequence of the covariance relations (\ref{eq:potentialCovar1})-(\ref{eq:LaplaceInv1}). Based on the expression for the DOSm in (\ref{eq:DefnDOSergodic1}), the covariance relations allow us to conclude that if $E \not \in \mathrm{supp}(n_v)$, then there exists $\epsilon > 0$, such that for $\mathbb{P}$-a.e. $\omega$, one has $\langle \delta_x, \chi_{(E- \epsilon, E + \epsilon)}(H_\omega) \delta_x  \rangle = 0$, for all $x \in \mathcal{V}$, which, together with (\ref{eq_spectrum_local_DOSoM}), produces the equality in (\ref{eq_asspectrumESO_DOSm}).

Combining (\ref{eq_contispectraHD_basic}) with, respectively, (\ref{eq_spectrum_local_DOSoM}) and (\ref{eq_spectrum_DOSoM}), we can thus summarize this discussion in the following proposition:

\begin{prop} \label{prop_contispectrumHD-det}
Given a deterministic Schr\"odinger operator on a graph $\mathbb{G}=(\mathcal{V}, \mathcal{E})$ as described in (\ref{eq_Schrodop}). For each $x \in \mathcal{V}$, the map $\mathit{l}^\infty(\mathbb{G}; \mathbb{R}) \ni V \mapsto \mathrm{supp}(n_{V;x}^*)$ is Lipschitz continuous in the Hausdorff metric, i.e. for $V,W \in \mathit{l}^\infty(\mathbb{G}; \mathbb{R})$:
\begin{equation}
\mathrm{dist}_{\rm H}( \mathrm{supp}(n_{V;x}^*) , \mathrm{supp}(n_{W;x}^*) ) \leq \Vert V - W \Vert_\infty ~\mbox{.}
\end{equation}
Similarly, the map $\mathit{l}^\infty(\mathbb{G}; \mathbb{R}) \ni V \mapsto \mathrm{supp}(n_{V}^*)$, satisfies
\begin{equation}
\mathrm{dist}_{\rm H}( \mathrm{supp}(n_{V}^*) , \mathrm{supp}(n_{W}^*) ) \leq \Vert V - W \Vert_\infty ~\mbox{.}
\end{equation}
\end{prop}


For ESO, Proposition \ref{prop_contispectrumHD-det} implies continuity of the a.s.-spectrum in the sampling function in $L^\infty$-norm. Considering RSO as a special case of ESO, where the role of the sampling function is played by the quantile function (\ref{eq:quantile1}) associated with the single-site probability measure, (\ref{eq_contispectrumHS_ESO}) yields continuity of the spectrum in the underlying single-site measure. We summarize these results in the following proposition.

\begin{prop}\label{prop:eso_rso_sp_cont1}
\begin{itemize}
\item[(i.)] For ESO  on a graph $\mathbb{G}=(\mathcal{V}, \mathcal{E})$ with underlying probability space $(\Omega, \mathcal{B}, \Pp)$, as described in section \ref{subsec:ergodicSO1}, the a.s.-spectrum $\Sigma_u$ is Lipschitz continuous with respect to the sampling function in the $ L^\infty$-norm. That is, for two sampling functions $u,v \in L^\infty ( (\Omega, \Pp); \R)$ taking values in $[-C,C]$, we have
\begin{equation} \label{eq_contispectrumHS_ESO}
\mathrm{dist}_{\rm H}(\Sigma_v , \Sigma_u) \leq \Vert v - u \Vert_\infty ~\mbox{.}
\end{equation}

\item[(ii.)]
For RSO, considered as a special case of ESO as described in section \ref{subsec:RSO1}, the a.s-spectrum $\Sigma_\mu$ is Lipschitz continuous with respect to the quantile function $q_\mu$, defined in (\ref{eq:quantile1}), associated with the single-site probability measure $\mu$. 
That is, for each $0<C<+\infty$, and two single-site probability measures $\mu,\nu \in \mathcal{P}([-C,C])$, one has
\begin{equation}  \label{eq_contispectrumHS_RSO}
\mathrm{dist}_{\rm H}(\Sigma_\mu , \Sigma_\nu) \leq \Vert q_\mu - q_\nu \Vert_{L^\infty ([0,1]; \mu_L)} ~\mbox{.}
\end{equation}
(See \eqref{eq_UBsupports} for an improved upper bound in terms of the supports of the single-site probability measures.)    
\end{itemize}
\end{prop}

Notice that the continuity of the map $\mu \mapsto \Sigma_\mu$ given in (\ref{eq_contispectrumHS_RSO}) equips the space of Borel probability measures $\mathcal{P}([-C,C])$ with the metric 
\begin{equation} \label{eq_Wassersteininfty}
d_\infty(\mu,\nu):=\Vert q_\mu - q_\nu \Vert_{L^\infty ([0,1]; \mu_L)} ~\mbox{.}
\end{equation}
In comparison, the continuity result for the DOSm in the single-site probability measure formulated in Theorem \ref{thm_ergodic_random_DOS} uses the weak topology on $\mathcal{P}([-C,C])$. Here, the weak topology is expressed by the Kantorovich-Rubinstein-Wasserstein metric $d_{\rm KRW}$ defined in (\ref{eq:KRWmetric1}). Recall, from (\ref{eq_equalitymetrics}), that the distance $d_{\rm KRW}$ is equivalent to the distance of the quantile functions in $L^1 ([0,1]; \mu_L)$; in particular, by (\ref{eq_equalitymetrics}), one has
\begin{equation} \label{eq_relationmetricsmeasures}
d_w \leq d_{\rm KRW} \leq d_\infty ~\mbox{.}
\end{equation}

We observe that metric in (\ref{eq_Wassersteininfty}) has a well-known equivalent formulation (see e.g. \cite[p. 650]{rachev}) as
\begin{equation} \label{eq_Wassersteininfty_equiv}
d_\infty(\mu,\nu) = \inf \{ \epsilon > 0 ~:~ \mu(A) \leq \nu(A^\epsilon) ~\mbox{ and } \nu(A) \leq \mu(A^\epsilon) ~\mbox{, for all $A \in \mathcal{B}$} \} ~\mbox{,}
\end{equation}
where, for $A \subseteq \mathbb{R}$, we let
\begin{equation}
A^\epsilon:=\{ x \in \mathbb{R} ~:~ \mathrm{dist}(x,A) < \epsilon \} ~\mbox{.}
\end{equation}
Since the Hausdorff distance between $\mathrm{supp}(\mu)$ and $\mathrm{supp}(\nu)$ can be characterized by
\begin{align} \label{eq_hd_equivdef}
\mathrm{dist}_{\rm H}( \mathrm{supp}(\mu) ,  \mathrm{supp}(\nu)  ) = \inf\{ \epsilon > 0 ~:~&  \mathrm{supp}(\mu) \subseteq (\mathrm{supp}(\nu))^\epsilon \nonumber \\
 &  ~\mbox{ and } \mathrm{supp}(\nu) \subseteq (\mathrm{supp}(\mu))^\epsilon \} ~\mbox{,}
\end{align}
(\ref{eq_Wassersteininfty_equiv}) shows that 
\begin{equation} \label{eq_distsuppWassersteininfty}
\mathrm{dist}_{\rm H}( \mathrm{supp}(\mu) ,  \mathrm{supp}(\nu)  ) \leq d_\infty(\mu,\nu) ~\mbox{.}
\end{equation}
Note that the inequality (\ref{eq_distsuppWassersteininfty}) is in general optimal since otherwise (\ref{eq_relationmetricsmeasures}) would imply that convergence of the support of measures in the Hausdorff metric implies weak convergence, which is false in general.

Finally, we mention that the quantitative continuity result in (\ref{eq_contispectrumHS_RSO}) can be improved using the well known expression for the a.s.-spectrum of random Sch\"odinger operators by Kunz-Souillard \cite{KunzSouillard}, later generalized by Kirsch-Martinelli \cite{KirschMartinelli_1982}.  For a RSO as in (\ref{eq:RSO1}) with a single-site probability measure $\mu \in \mathcal{P}(\mathbb{R})$, the a.s.-spectrum is given by
\begin{equation} \label{eq_RSO_spectrumKunzSouillard}
\Sigma_\mu = \sigma(\Delta_\mathbb{G}) + \mathrm{supp}(\mu) ~\mbox{.}
\end{equation}
Thus, the equality (\ref{eq_RSO_spectrumKunzSouillard}) improves (\ref{eq_Wassersteininfty}) to
\begin{align} \label{eq_UBsupports}
\mathrm{dist}_{\rm H}(\Sigma_\mu , \Sigma_\nu) & = \mathrm{dist}_{\rm H}(\sigma(\Delta_\mathbb{G}) + \mathrm{supp}(\mu) , \sigma(\Delta_\mathbb{G}) + \mathrm{supp}(\nu) ) \nonumber \\
 & \leq \mathrm{dist}_{\rm H}( \mathrm{supp}(\mu) ,  \mathrm{supp}(\nu)  ) ~\mbox{.}
\end{align}

We conclude with the following example based on (\ref{eq_RSO_spectrumKunzSouillard}) which, in view of our main result for RSO in Theorem \ref{thm_ergodic_random_DOS},  illustrates that the map $\mu \mapsto \Sigma_\mu$ is {\em{not}} continuous  when equipping the domain $\mathcal{P}([-C,C])$ with the weak topology (and the codomain with the Hausdorff metric as before):

\begin{example} \label{example}
For $n \in \mathbb{N}$, consider the RSO on $\mathbb{Z}^d$ with underlying single-site (Bernoulli) measure
\begin{equation}
\mu_n:= (1- \frac{1}{n}) \delta_0 + \frac{1}{n} \delta_{100d} ~\mbox{.}
\end{equation}
On the one hand, the sequence $(\mu_n)$ converges weakly with
\begin{equation}
\mu_n \stackrel{w}{\to} \delta_0 =: \mu ~\mbox{.}
\end{equation}
In particular, Theorem \ref{thm_ergodic_random_DOS} (i), implies convergence of the DOSm with
\begin{equation}
d_w(n_{\mu_n} , n_\mu) \leq d_{\rm KRW}(\mu_n, \mu) = 100d \cdot \frac{1}{n} \to 0 ~\mbox{.}
\end{equation}
On the other hand, using (\ref{eq_RSO_spectrumKunzSouillard}), we have
\begin{equation} \label{eq_example_concl}
\mathrm{dist}_{\rm H} (\Sigma_{\mu_n} , \Sigma_\mu) = 96 d ~\mbox{, for all $n \in \mathbb{N}$.}
\end{equation}

We mention that (\ref{eq_example_concl}) also provides an explicit example that the last inequality in (\ref{eq_UBsupports}) cannot, in general, be improved to an equality since $\mathrm{dist}_{\rm H}( \mathrm{supp}(  \mu_n) , \mathrm{supp}( \mu ) ) = 100 d$.
\end{example}

\end{appendices}


\end{document}